\documentclass[11pt,draftcls,onecolumn]{IEEEtran}
\usepackage{amsmath}
\usepackage{txfonts}
\usepackage{setspace}
\usepackage{enumerate}

\usepackage{stmaryrd}
\usepackage{amssymb}
\usepackage{mathrsfs}
\usepackage{amsfonts}  % include bezier curves
\usepackage{flushend}
\usepackage[dvips,final]{graphicx}
\usepackage[dvips]{color}
\usepackage{epsfig,amssymb}
\newcommand{\beq}{\begin{equation}}
\newcommand{\eeq}{\end{equation}}
\newcommand{\beqn}{\begin{eqnarray}}
\newcommand{\eeqn}{\end{eqnarray}}

\begin{document}

\begin{center}

{\Large\bf Measurement Matrix Design for Compressive Sensing Based
MIMO Radar \footnote{ This work was supported by the by the Office
of Naval Research under Grants ONR-N-00014-07-1-0500 and
ONR-N-00014-09-1-0342 and the National Science Foundation under
Grants CNS-09-05398 and CNS-04-35052 }}

\bigskip
{\it Yao Yu } \\
         Department of Electrical \& Computer Engineering,
         Drexel University, Philadelphia, PA 19104\\
         \medskip
         {\it Athina P. Petropulu}\\
         Department of Electrical \& Computer Engineering,
Rutgers, The State University of New Jersey, Piscataway, NJ
08854-8058
\\
         \medskip
         {\it H. Vincent Poor}\\
         School of Engineering and Applied Science,
         Princeton University, Princeton, NJ 08544
\end{center}

\bigskip

\begin{abstract}

In colocated multiple-input multiple-output (MIMO) radar using
compressive sensing (CS), a receive node  compresses its received
signal via a linear transformation, referred to  as  measurement
matrix. The samples are subsequently forwarded  to a fusion center,
where an
 $\ell_1$-optimization problem is formulated and solved for target
 information. CS-based MIMO radar exploits the target sparsity
in the angle-Doppler-range space and thus achieves the high
localization performance of traditional MIMO radar but with many
fewer measurements. The measurement matrix is vital for CS recovery
performance. This paper considers the design of measurement matrices
that achieve an optimality criterion that depends on the
 coherence of the sensing matrix (CSM) and/or
signal-to-interference ratio (SIR). The first approach minimizes a
performance penalty that is a linear combination of CSM and the
inverse  SIR. The second one imposes a structure on the measurement
matrix and determines the parameters involved so that the SIR is
enhanced. Depending on the transmit waveforms, the second approach
can significantly improve SIR, while maintaining CSM
 comparable to that of the
Gaussian random measurement matrix (GRMM).
 Simulations indicate that the
proposed measurement matrices can improve  detection accuracy as
compared to a GRMM.


\end{abstract}
\section{Introduction}

 Multiple-input
multiple-output (MIMO) radar has received considerable recent
attention \cite{Fishler:04}-\cite{Li:07}. A MIMO radar consists of
multiple transmit and receive antennas and is advantageous in two
different scenarios \cite{Haimovich: 08}-\cite{Chen:081}, namely,
{widely separated antennas} and {collocated antennas}. In the first
scenario \cite{Haimovich: 08}, the transmit antennas are located far
apart from each other relative to their distance to the target. The
MIMO radar system transmits independent probing signals from its
antennas that follow independent paths, and thus each target return
carries independent information about the target. Joint processing
of the target returns results in diversity gain, which enables the
MIMO radar to achieve high target resolution. Widely distributed
MIMO radar systems are shown to offer considerable advantages for
estimation of  target parameters, such as location
{\cite{p1-Godrich-Haimovich-2010}} and velocity
{\cite{p1-He-Blum-2010}}. In the collocated scenario \cite{Stoica:
07m}\cite{Chen:081}, the transmit and receive antennas are located
close to each other relative to the target, so that all antennas
view the same aspect of the target. In this scenario, the phase
differences induced by transmit and receive antennas can be
exploited to form a long virtual array with the number of elements
equal to the product of the numbers of transmit and receive nodes.
This enables the MIMO radar  to achieve superior resolution in terms
of direction of arrival (DOA) estimation and parameter
identification \cite{Stoica: 07m}.

Compressive sensing (CS) theory \cite{Donoho:06}-\cite{Romberg:08}
states that a signal $\mathbf{x}$ that exhibits sparsity in some
domain, can be recovered from a number of samples that is much
smaller than that required by  Nyquist theory. In particular, a
signal of length $N$ that can be represented by $K$ ($K<<N$) basis
vectors in some space, can be recovered
exactly with high probability from $\mathcal{O}(K\log N)$ measurements. Let $%
\mathbf{\Psi}$ denote the basis matrix that spans that space, and $\mathbf{%
\Phi}$ denote an $M\times N$ matrix with $M\ll N$, that is incoherent with $%
\mathbf{\Psi}$ and is referred to as the measurement matrix. The recovery
proceeds by finding the coefficients of the $K$ basis vectors in the signal
decomposition. This is formulated as an $\ell_1$-optimization problem, i.e.,
$\min\|\mathbf{s}\|_1,\ \ s.t. \ \text{to}\ \mathbf{y}=\mathbf{\Phi}\mathbf{x%
} =\mathbf{\Phi \Psi} \mathbf{s}.$ Throughput this paper, we will
refer to recovery along these lines   as the {\it CS approach}. The
product $\mathbf{\Phi\Psi}$ is usually referred to as the {\it
sensing matrix}. According to the uniform uncertainty principle
(UUP) \cite{Candes:08}-\cite{Candes:062}, if every set of  sensing
matrix columns with cardinality less than the sparsity of the signal
of interest is approximately orthogonal, then the sparse signal can
be exactly recovered with high probability. In other words, CS
recovery requires that  $\mathbf{\Phi}$ is incoherent with
$\mathbf{\Psi}$. For an orthonormal basis matrix,  use of a random
 measurement matrix leads to a sensing matrix that obeys the
UUP  with overwhelming probability \cite{Candes:061}. The entries of
such a measurement matrix can be taken from a Gaussian distribution
or symmetric Bernoulli distribution.  The rows of a Fourier matrix
or an orthonormal matrix could also compose a measurement matrix.
 In this paper, we term
as {\it the conventional approach}  CS recovery using a Gaussian
measurement matrix.

The application of  CS to radar and MIMO radar  has been explored in
\cite{Baraniuk:07}-\cite{Herman:08} and
\cite{Petropulu:08}-\cite{Yu:09_tsp}, respectively. In both
\cite{Chen:08} and \cite{Strohmer:Asilomar09}, the authors
considered a uniform linear array  as a transmit and receive antenna
configuration and proposed to use a submatrix of the identity matrix
as the measurement matrix. %The proposed CS approaches in
%\cite{Chen:08}\cite{Strohmer:Asilomar09} are easily implementable in
%hardware, but they are  sensitive to interference.
In \cite{Petropulu:08} and \cite{Yu:09_tsp},  a CS-based MIMO radar system implemented on a small scale
network was proposed.
The network consists of a number of transmit and
receive nodes, each equipped with a single antenna, that are randomly
distributed over a small area. Each transmit node transmits a different narrowband signal.
 If the number of targets is small, the signal that is
reflected by  targets and is picked up at a receive node is sparse
in the angle-range-Doppler space. This fact can be exploited to
achieve target detection and localization using only a small number
of compressively obtained samples at each receive node, and/or by
involving a small number of receive nodes \cite{Yu:09_tsp}. The
approach of \cite{Petropulu:08} and \cite{Yu:09_tsp} was applied to
 the case in which the
targets are located within a small range bin and the sampling is
synchronized with the first target return.
To improve performance  in the
presence of strong interference
the columns of the sensing matrix were
designed
 to incorporate
information on the transmit waveforms.

In this paper, we consider a general scenario that does not confine
the targets within a small range bin, nor does it require sampling
synchronization. When the targets are separated by several range
bins, different targets will introduce different delays in the
received waveforms. In that case, the formulation of
\cite{Petropulu:08} and \cite{Yu:09_tsp} no longer applies. This
problem was considered  in \cite{Yu:10_aes}, where a step-frequency
approach was proposed in order to improve range resolution. Here,
our goal is optimal or suboptimal measurement matrix design that
decreases the coherence of the sensing matrix (CSM) and/or enhances
signal-to-interference ratio (SIR). The first design minimizes a
performance penalty that is a linear combination of CSM and the
inverse SIR. The measurement matrix is obtained by solving a convex
optimization problem that involves high computational complexity.
 A suboptimal solution is also proposed that forces a specific structure to the measurement matrix.
The second design
 targets only SIR improvement; it is constructed based on the transmit
signal waveforms and accounts for all possible discretized delays of
target returns within a given time window. It is shown that
depending on the waveforms used, the  latter measurement matrix can
significantly improve SIR while it results in  CSM  comparable to that
of  the random Gaussian measurement matrix.

 The rest of the paper  is organized as follows.  In Section II we provide
the signal model of a CS-based MIMO radar system with targets
falling in different range bins. In Section III, we introduce the
two proposed  measurement matrices and provide the SIR analysis
related to the second measurement matrix.  Simulation results are
given in Section IV for stationary targets. Finally, we make some
concluding remarks in Section V.

\emph{Notation}: Lower case and capital letters in bold denote
respectively vectors and matrices.  The  expectation of a random
variable is denoted by $E\{\cdot\}$. Superscripts
 $ (\cdot)^{H}$ and $\mathrm{Tr}(\cdot)$ denote respectively the
Hermitian transpose and  trace  of a matrix. $A(m,n)$ represents the
$(m,n)$-th entry of the matrix $\bf A$. ${\mathbf 0}_{L\times M}$
denotes an $L\times M$ matrix with zero entries.

\section{Signal Model for CS-based  MIMO Radar}\label{sig_model}
Let us consider a MIMO radar system consisting  of  $M_t$ transmit
antennas (TXs) and $N_r$ receive antennas (RXs) that are randomly
distributed over a small area (colocated).  Each TX node transmits
periodic narrowband pulses of duration $T_p$ and pulse repetition
interval (PRI)
 $T$.
 Let $(r^t_{i}, \alpha^t_{i})$/$(r^r_{i},
\alpha^r_{i})$   denote the location of the $i$-th transmit/receive
node   in polar coordinates.
Let us also consider the presence of
 $K$ slowly-moving point targets located in different range bins; the $k$-th target is at azimuth angle $\theta_k$ and moves
with constant radial speed $v_k$.

Let $d_k(t)$ denote the range of the $k$-th target at time $t$.
 Under the far-field assumption, i.e.,
 $d_{k}(t) \gg r^{t/r}_{i}$, the distance between the $i$th transmit/receive
node  and the $k$-th target
 $d^t_{ik}$/$d^r_{ik}$ can be approximated as
\begin{eqnarray}
d^{t/r}_{ik}(t) \approx d_k(t)- \eta_i^{t/r}(\theta_k)
=d_k(0)-\eta_{i}^{t/r}(\theta_k)-v_kt
\end{eqnarray}
where
$\eta_{i}^{t/r}(\theta_k)=r^{t/r}_{i}\cos(\theta_k-\alpha^{t/r}_{i})$.
 Let us consider the return from the $k$-th target arriving at the $l$-th
antenna during the $m$-th pulse, i.e.,
\begin{align} \label{rec_sig_1}
y^k_{lm}(t)&= \sum_{i=1}^{M_t}\beta_k
x_i(t-(d^{t}_{ik}(t)+d^{r}_{lk}(t))/c)
\exp({j{2\pi}f(t-(d^t_{ik}(t)+d^{r}_{lk}(t))/{c}) })
\end{align}
where $c$, $f$  and $\beta_{k}$  denotes  the speed of light, the
carrier frequency, and the reflection coefficient of the $k$-th
target, respectively;  $x_i(t)$ represents  the transmit waveform of
the $i$-th node. Under the narrowband assumption, and due to the
slow target speed so that the Doppler shift is negligible, the
baseband signal corresponding to (\ref{rec_sig_1}) becomes
\begin{align} \label{rec_sig}
y^k_{lm}(t)& \approx \sum_{i=1}^{M_t}\beta_k x_i(t-2d_k(0)/c)
\exp(-{j{2\pi}f(d^t_{ik}(t)+d^{r}_{lk}(t))/{c} }).
\end{align}
Due to the closeness of the transmit and receive nodes, the
distances between nodes and the target are approximately the same
for all receivers. Thus, the time delay in the waveforms, induced by
the $k$-th target can be approximately based on the range
corresponding to  the initial sampling time, i.e., $d_k(0)$, which
is independent of the RX index. The $l$-th node compressively
samples the return signal  to obtain $M$ samples per pulse (please
refer to Fig. 1 of \cite{Yu:09_tsp} for a schematic of the
receiver). Let $L$ denote the number of $T_p/L$-spaced samples of
the transmitted waveforms within one pulse. The effect of the
compressive receiver of Fig. 1 of \cite{Yu:09_tsp} is equivalent to
pre-multiplying by matrix $\mathbf{\Phi}_l$
 a $T_p/L$-sampled version of the received pulse. The size of $\mathbf{\Phi}_l$ is $M \times (L+\tilde L)$, where $\tilde L$ is the maximum
delay among the return signals normalized by $T_p/L$ and is known in
advance. Here $M<<L$. The obtained samples are then placed in vector
${\mathbf{r}}_{lm}$, which can be expressed in matrix form as
\cite{Yu:10_aes}
\begin{align}\label{rec_sig}
{{\bf r}}_{lm}&= \sum_{k=1}^{K}\beta_ke^{j2\pi {p}_{lmk}}{\bf
\Phi}_l{\bf D}(f_{k})\mathbf{C}_{\tau_k}{\bf X}{\bf
v}(\theta_k)+{\bf \Phi}_l{\bf n}_{lm}
\end{align}
where
\begin{enumerate}
\item ${p}_{lmk}=\frac{-2d_k(0)
f}{c}+\frac{\eta_{l}^{r}(\theta_k)f}{c}+{f_{k}(m-1)T}$, where
$f_{k}=\frac{2v_k f}{c}$ is the Doppler shift induced by the $k$-th
target; $\bf X$ is an $L\times M_t$ matrix that contains the
transmit waveforms of $M_t$ antennas as its columns and
$\mathrm{diag}\{{\bf X}^H{\bf X}\}=[1,\ldots,1]^T$;

\item ${\mathbf{\Phi}}_l$
 is the $M\times (L+\tilde{L})$ measurement matrix for the $l$-th receive
node;
\item ${\bf v}(\theta_k)=[e^{j\frac{2\pi
f}{c}\eta^t_{1}(\theta_k)},...,e^{j\frac{2\pi
f}{c}\eta^t_{M_t}(\theta_k)}]^T$
 and ${\bf D}(f_{k})={\rm
diag}\{[e^{j{2\pi}f_{k}0T_p/L},\ldots,e^{j{2\pi}f_{k}(L+\tilde{L}-1)T_p/L}]\}$;
\item $\tau_k=\lfloor\frac{2d_k(0)}{cT_p/L}\rfloor$ and ${\bf
C}_{\tau_k}=[{\bf 0}_{L\times \tau_k},{\bf I}_{L},{\bf 0}_{L\times
(\tilde{L}-\tau_k)}]^T$. Here, we assume that the target returns
completely fall within the sampling window of length
$(L+\tilde{L})T_p/L$, and that $T_p/L$ is small enough so that the
rounding error in the delay is small, i.e., $x_i(t-\tau_k)\approx
x_i(t-\frac{2d_k(0)}{cT_p/L})$.
\item ${\bf n}_{lm}$ is
the interference of variance $\sigma^2$ at the $l$-th receiver
during the $m$-th pulse, arising due to the jammer signals and
thermal noise.
\end{enumerate}
 Discretize the angle, speed and range
space on a fine grid, i.e., respectively,
$[\tilde{a}_1,\ldots,\tilde{a}_{N_a}]$,
$[\tilde{b}_1,\ldots,\tilde{b}_{N_b}]$ and
$[\tilde{c}_1,\ldots,\tilde{c}_{N_c}]$. Let the grid points be
arranged first angle-wise, then range-wise,  and finally speed-wise
to yield the grid points $(a_n,b_n,c_n),
 n=1,...,N_a N_b N_c$.
 Through this ordering,   the grid point
$(\tilde{a}_{n_a},\tilde{b}_{n_b},\tilde{c}_{n_c})$ is mapped to
point
  $(a_n,b_n,c_n)$  with
$n=(n_b-1)n_an_c+(n_c-1)n_a+n_a$.
%\begin{align}
%n_c=\lfloor \frac{n-1}{N_aN_b}\rfloor+1,\
%n_b=\lfloor\frac{n-(n_c-1)N_aN_b-1}{N_a}\rfloor+1,\ n_a=
%n-(n_c-1)N_aN_b-(n_b-1)N_b.
%\end{align}
The discretized step is small enough so that each target falls on
some angle-speed-range
 grid point. Then (\ref{rec_sig})   can be
rewritten as
\begin{align}\label{received signal}
{{\bf r}}_{lm}&={\bf \Phi}_l\left( \sum_{n=1}^{N}s_ne^{j2\pi
{q}_{lmn}}{\bf
D}\left(\frac{2b_nf}{c}\right)\mathbf{C}_{\lfloor\frac{2c_n}{cT_p/L}\rfloor}{
\bf X}{\bf v}(a_n)+{\bf n}_{lm}\right)
\end{align}
where $N=N_aN_bN_c$, $
 s_n = \left\{
\begin{array}{rl}
\beta_{k},  &  \text{if the $k$-th target is  at}\ (a_n,b_n,c_n) \\
0,  & \text{otherwise}
\end{array} \right.$ and
\begin{align}
q_{lmn}=\frac{-2c_n
f}{c}+\frac{\eta_{l}^{r}(a_n)f}{c}+\frac{2b_nf(m-1)T}{c}.
\end{align}
 In  matrix form we have $ {{\mathbf
r}}_{lm}=\mathbf{\Theta}_{lm}{\mathbf{s}}+{\bf \Phi}_l{\bf n}_{lm}
$, where  ${\bf s}=[s_1,...,s_N]^T$ and
\begin{align}\label{sensing_matrix}
\mathbf{\Theta}_{lm}={{\bf \Phi}}_l\underbrace{[e^{j2\pi
q_{lm1}}{\bf
D}(2b_1f/c)\mathbf{C}_{\lfloor\frac{2c_1}{cT_p/L}\rfloor}{{\bf
X}}{\bf v}(a_1),\ldots,e^{j2\pi q_{lmN}}{\bf
D}(2b_Nf/c)\mathbf{C}_{\lfloor\frac{2c_N}{cT_p/L}\rfloor}{{\bf
X}}{\bf v}(a_N)]}_{\mathbf{\Psi}_{lm}}.
\end{align}
According to the CS formulation, $\mathbf{\Theta}_{lm}$ is the
sensing matrix
 and $\mathbf{\Psi}_{lm}$ is the  basis matrix.

If the number of targets is small as compared to $N$,  then the
positions of the targets are sparse in the angle-speed-range space
and $\mathbf{s}$ is a sparse vector. The locations of the non-zero
elements of ${\bf s}$ provide information on target angle, speed and
range. All the receive nodes forward their compressed measurements
to a fusion center. We assume that the fusion center has the ability
to separate the data of different nodes from each other. This can be
done,  for instance,  if the nodes send their data over different
carriers. The fusion center
 combines the compressively sampled signals due to $N_p$ pulses
obtained at $N_r$ receive nodes to form the vector  ${{\bf r}}$. Using the predefined measurement matrices,
 the discretization of the
angle-speed-range space, and also knowledge of the waveform matrix
${\bf X}$, the fusion center  obtains an estimate of  $\mathbf{s}$
by applying the Dantzig selector \cite{Candes:07}.

\section{Measurement matrix design }\label{measurment_matrix_design}
In this section, we discuss the design of the measurement matrix in
order to improve the detection performance of CS-MIMO radar. For the
sake of simplicity, we assume that all the nodes use the same
measurement matrix, denoted by ${\bf \Phi}$,  which does not vary
with time. Since the targets are moving slowly, the Doppler shift
within a pulse can be ignored. Generally, there are two factors that
affect  the performance of CS. The first one is the coherence of the
sensing matrix. UUP requires low CSM to guarantee exact recovery of
the sparse signal. The second factor is SIR. If the basis matrix
obeys the UUP and the signal of interest $\bf s$ is sufficiently
sparse, then the square estimation error of the Dantzig selector
satisfies with very high probability \cite{Candes:07}
\begin{eqnarray}\label{error_bound}
\parallel\hat{\bf s}-{\bf s}\parallel_{\ell_2}^2 \leq C^2 2log
N\times\left({\sigma}^2+\sum_{i}^{N}\min(s^2(i),{{\sigma}}^2)\right)
\end{eqnarray}
where $C$ is a constant.  It can be easily seen from
(\ref{error_bound}) that an increase in the interference power
degrades the performance of the Dantzig selector.

\subsection{Measurement matrix design $\#1$}\label{measurement_matrix_opti}

 The goal of measurement matrix design is to reduce the
coherence of the sensing matrix and at the same time increase
SIR.
 The coherence of two columns of
the sensing matrix, ${\bf \Theta}$, corresponding to the $k$-th and
$k'$-th grid point is given by
\begin{align}
\mu_{kk'}({\mathbf{\Theta}})&=\frac{\left
|\sum_{m=1}^{N_p}\sum_{l=1}^{N_r}e^{j2\pi(q_{lmk}-q_{lmk'})}\left
({\bf \Phi}{\bf C}_{\lfloor \frac{2c_{k'}}{cT_p/L}\rfloor}{\bf
X}{\bf v}(a_{k'})\right)^H{\bf \Phi}{\bf C}_{\lfloor
\frac{2c_{k}}{cT_p/L}\rfloor}{\bf X}{\bf v}(a_{k})  \right
|}{N_r\sqrt{\sum_{m=1}^{N_p}\left\|{\bf \Phi}{\bf C}_{\lfloor
\frac{2c_{k}}{cT_p/L}\rfloor}{\bf X}{\bf v}(a_{k})\right\|^2_2
\sum_{m=1}^{N_p}\left\|{\bf \Phi}{\bf C}_{\lfloor
\frac{2c_{k'}}{cT_p/L}\rfloor}{\bf X}{\bf v}(a_{k'})\right\|^2_2
}}\nonumber\\
&=\frac{\left
|\sum_{m=1}^{N_p}\sum_{l=1}^{N_r}e^{j2\pi(q_{lmk}-q_{lmk'})} {\bf
u}^H_{k'}{\bf \Phi}^H{\bf \Phi}{{\bf u}}_{k} \right |}{
N_rN_p\sqrt{{\bf u}^H_{k}{{\bf \Phi}}^H{\bf \Phi}{{\bf u}}_{k}{\bf
u}^H_{k'}{{\bf \Phi}}^H{\bf \Phi}{{\bf u}}_{k'} }}
\end{align}
where ${\bf u}_{k}={\bf C}_{\lfloor
\frac{2c_{k}}{cT_p/L}\rfloor}{\bf X}{\bf v}(a_{k})$.

Let the interference waveform at the $l$-th receive node during the
$m$-th pulse be Gaussian distributed, i.e.,
${n}_{lm}(t)\sim\mathcal{CN}(0,\sigma^2)$. Let us  also assume that
the noise waveforms are independent across receive nodes and between
pulses. Then the average power of the interference equals
\begin{align}\label{inter_power}
 P_n&=E\left\{\sum_{m=1}^{N_p}\sum_{l=1}^{N_r}({\bf \Phi}{{\bf n}}_{lm})^H{\bf
\Phi}{{\bf n}}_{lm}\right\}=N_pN_r\sigma^2\mathrm{Tr}\{{\bf
\Phi}^H{{\bf \Phi
 }}\}.
\end{align}
The average power of the  echo reflected by the   $i$-th target
located on the $k_i$-th grid point of the angle-range space is
approximately equal to
\begin{align}
 P^i_s
 %&=E\{({\bf r}^k_{lm})^H{\bf r}^k_{lm} | {\bf X}\}\nonumber\\
&\approx |\beta_i|^2 N_rN_p{\bf u}^H_{k_i}{{\bf \Phi}}^H{\bf
\Phi}{{\bf u}}_{k_i}.
\end{align}
Therefore, the SIR equals approximately
\begin{eqnarray}\label{measure_opt}
\mathrm{SIR}\approx\frac{\sum_{i=1}^K|\beta_i|^2 {\bf u}^H_{k_i}{\bf
\Phi}^H{{\bf \Phi}}{{\bf u}}_{k_i} }{\sigma^2\mathrm{Tr}\{{{\bf
\Phi}}^H{{\bf \Phi }}\}}.
\end{eqnarray}

The precise manner in which  CSM and  SIR affect the performance of
the CS approach is unknown. Although  theoretical bounds for the
$\ell_2$-norm of the estimation error have been proposed
\cite{Candes:07}-\cite{Needell:09}, those bounds  might not be
relevant in applications in which the quantities of interests are
the locations of the non-zero elements of the sparse signal, rather
than the non-zero values themselves. This is the case in the problem
at hand.  In \cite{Tang:09},  an upper bound on the error
probability of sparse support recovery, i.e., the total probability
of missed detection and false alarm,  under the {optimal} decision
rule was derived. Although that upper bound is related to the
detection of non-zero elements, it cannot be used for the design of
the measurement matrix because it is rather loose, and further, it
involves eigenvalues of submatrices of a given sensing matrix
corresponding to all possible sparse patterns for the signal of
interest.
%[YAO: I do not understand the following and how \cite{Chen:Asilomar08}  relates to \cite{Tang:09}]
%Since the entries of the measurement matrix can take
%any values, one cannot even use the heuristic
%approaches of \cite{Chen:Asilomar08} to obtain the measurement
%matrix from the bound of \cite{Tang:09}.

 In this paper, we
determine the measurement matrix by optimizing a linear combination
of  CSM and the reciprocal
 of SIR. The  CSM  can be defined in various ways.
Let us define CSM  as
 the maximum coherence produced by a pair of cross columns in the sensing
matrix.  This criterion works well
 for a uniform sensing matrix
but  might not  capture the behavior of  the sensing matrix in cases
in which the coherence of most column pairs  is small
\cite{Tropp:04}. However, this coherence metric is widely used for
the CS scenario due to its simplicity
\cite{Tropp:04}\cite{Tropp:09}.

The
optimization problem becomes
\begin{eqnarray}
\min_{{{\bf \Phi}}}\left(\max_{k\neq
k'}\mu^2_{kk'}({\mathbf{\Theta}})+\lambda
\frac{1}{\mathrm{SIR}}\right)\ \
\end{eqnarray}
where $\lambda$ is a positive weight, which reflects the tradeoff
between the CSM and SIR.

The problem of (\ref{measure_opt}) is not convex. In order to obtain
a solution, let us first view (\ref{measure_opt}) as an optimization
problem with respect to ${\bf B}={{\bf \Phi}}^H{{\bf \Phi}}$.
Furthermore, let us set  the norm of the columns of  the sensing
matrix  to $1$, i.e., $N_rN_p{\bf u}^H_{k}{{\bf \Phi}}^H{{\bf
\Phi}}{{\bf u}}_{k}=1,\ k=1,...,N$;  this will significantly
simplify the  expression for $\mu_{kk'}(\mathbf{\Theta})$ and
$\frac{1}{\mathrm{SIR}}$.
 Now, (\ref{measure_opt}) can be
reformulated (using the approximation (\ref{measure_opt})) as
\begin{eqnarray}\label{measure_opt_new}
\min_{t, {\bf B}}&&t+\lambda\mathrm{Tr}\{{\bf B}\}\nonumber\\
 s.t. &&\left|\sum_{m=1}^{N_p}\sum_{l=1}^{N_r}e^{j2\pi(q_{lmk'}-q_{lmk})}
{\bf u}_{k'}^H{\bf B}{\bf u}_{k} \right |^2\leq
t,\nonumber\\
&&k=1,\ldots, N, k'=k+1,\ldots, N\nonumber\\
&&   N_rN_p\mathbf{u}^H_{k}{\bf B}\mathbf{u}_{k}=1, \
k=1,\ldots, N,\nonumber\\
&& {\bf B}\geq 0,\ t\geq 0,
\end{eqnarray}
which  is a convex problem with respect to $\bf B$. The first term
in the objective refers to the maximum coherence of cross columns in
the sensing matrix; the second term is proportional to the noise
power which is a linear function of $\bf B$. Once $\bf B$ is
obtained, the solution of (\ref{measure_opt})
 can be  obtained based on  the
eigendecomposition $\bf B={\bf V}\Sigma {\bf V}^H$, as
\begin{align}
{\bf \Phi}_{\#1}=\sqrt{\tilde{\bf \Sigma}} \tilde{{\bf V}}^H
\end{align}
 where $\tilde{\bf \Sigma}$ is a diagonal matrix that contains on its diagonal
the nonzero eigenvalues of ${\bf \Sigma}$,  and $\tilde{{\bf V}}$
contains as its columns the corresponding eigenvectors.

 Another
definition for CSM would be as the sum of the coherence of all pairs
of columns in the sensing matrix (SCSM). With this measure of CSM,
the minimization problem becomes
\begin{eqnarray}\label{measure_opt2}
\min_{{{\bf \Phi}}}\left(\sum_{k\neq
k'}\mu^2_{kk'}({\mathbf{\Theta}})+\lambda
\frac{1}{\mathrm{SIR}}\right).
\end{eqnarray}
Simulation results show that the solution of (\ref{measure_opt2})
can increase the number of column pairs in the sensing matrix that
have low coherence as compared to the solution of
(\ref{measure_opt_new}). The latter solution
 tends to increase the coherence of some column pairs
in the sensing matrix which had low coherence. This is intuitively expected  because
solving (\ref{measure_opt_new}) requires more constraints than (\ref{measure_opt2}).
In this paper,  we use SCSM to design the measurement matrix.

 The above proposed methods for optimizing the
measurement matrix  reduce the coherence of cross columns in the
sensing matrix without amplifying the interference. These methods
improve the detection performance of the CS-based MIMO radar system,
but incur an increased computational load as compared to CS-based
MIMO radar that relies on the conventional measurement matrix. The
number of complex variables entering  the convex problem of
(\ref{measure_opt2}) is $\frac{(\tilde{L}+L+1)(\tilde{L}+L)}{2}$.
The computation complexity would be  prohibitively high for large
values of $\tilde{L}+L$. Also, for a large number of grid points
$N$, we have to deal with a large number of constraints. The optimal
measurement matrix might be obtained and stored offline based on
knowledge of grid points in the angle-range space. However, it would
need to be updated once the basis matrix changes with the search
area of interest. This would bring heavy burden to radar systems and
thus might render the real-time application impossible. Therefore,
ways to alleviate the computational load are worthy of
investigation. A suboptimal scheme for this problem that involves
lower complexity is discussed next.

Let us impose a  structure on the measurement matrix to be
determined  as follows:
\begin{align}\label{subopt1}
{{\bf \Phi}}_{\#1}={\bf W}{\bf \Phi}
\end{align}
where ${\bf W}$ is an $(L+\tilde{L})\times \tilde{M} $ unknown
matrix to be determined and ${\bf \Phi}$ is a $\tilde{M}\times
M_t(\tilde{L}+1)$  Gaussian random matrix. Then the number of
variables in ${\bf W}$ can be controlled by changing the value of
$\tilde{M}$. We can obtain ${\bf W}$ by solving (\ref{measure_opt2})
with ${\bf B}={{\bf W}^H{\bf W}}$ and ${\bf u}_{k}={\bf \Phi}{\bf
C}_{\lfloor \frac{2c_{k}}{cT_p/L}\rfloor}{\bf X}{\bf v}(a_{k})$.
Furthermore, the structure in (\ref{subopt1}) enables a two-step
processing for CS-based MIMO radar that simplifies the hardware of
the receive nodes. In particular, a receive node linearly compresses
the incident signal by using ${\bf \Phi}$. At the fusion center, all
the signals forwarded by receive nodes are first multiplied by $\bf
W$ and then jointly processed to extract target information. We can
think of $\bf W$ as a type of post processing. In this way, the
received nodes require no information about $\bf W$, which reduces
the communication overhead for the fusion center and nodes.

In order to render the convex problem tractable, the norm of the
columns in the sensing matrix is set to be a constant. This
increases the number of constraints. If the number of variables is
not sufficiently high, there might not be enough degrees of freedoms
to decrease the coherence of the sensing matrix as compared to the
original one. Since  the number of constraints equals the number of
grid points, the number of constraints  can be decreased by reducing
the search area.  This can be done by considering grid points around
some initial angle-range estimates, if such estimates are available.

\subsection{Measurement  matrix  design $\#2$} \label{measurement_matrix_subopti}

  Although the
suboptimal construction in (\ref{subopt1}) significantly reduces the
number of variables, solving (\ref{measure_opt2}) still requires
high computational load. Further, the solution
  needs to be adapted to a
particular basis matrix. In order to avoid these two shortcomings
of  ${{\bf \Phi}}_{\#1}$, we next propose another  measurement
matrix that targets SIR improvement only.

 As in
\cite{Yu:09_tsp}, we impose a special structure on the measurement
matrix, i.e.,
\begin{equation}
{\bf \Phi}_{\#2}={\bf \Phi W}^H \ , \end{equation} where
$\mathbf{\Phi}$ is an $M\times \tilde{M} \ (M\leq \tilde{M})$
zero-mean Gaussian random matrix and ${\bf W}$ is an $(L+\tilde
L)\times \tilde{M}$ deterministic matrix satisfying
$\mathrm{diag}\{{\bf W}^H{\bf W}\}=[1,\ldots,1]^T$.  The above
structure serves two purposes. First, the matrix  ${\bf W}$ can be
selected to improve the detection performance of the CS approach  at
the receiver. Second, ${\bf \Phi}_{\#2}$ is always  Gaussian
regardless of ${\bf W}$.     As will be shown next,    with the
appropriate  ${\bf W}$,  ${\bf \Phi}_{\#2}$
 may not
result is higher CSM as compared to the conventional measurement
matrix.
Next, we discuss the selection of ${\bf W}$.

The average power of the  echo reflected by the  $k$-th target  with
respect to node locations,
 conditioned on the transmit waveforms, is
approximately equal to
\begin{align}\label{sig_power_k}
 P^k_s
 %&=E\{({\bf r}^k_{lm})^H{\bf r}^k_{lm} | {\bf X}\}\nonumber\\
&\approx  |\beta_k|^2N_rN_pE\{\mathrm{Tr}\{{{\bf \Phi }}{\bf
W}^H{\bf C}_{\tau_k}{\bf X}{\bf v}(\theta_k)({{\bf \Phi }}{\bf
W}^H{\bf C}_{\tau_k}{\bf X}{\bf v}(\theta_k))^H\}\}
\nonumber\\&\approx |\beta_k|^2N_rN_p\mathrm{Tr}\{{{\bf \Phi}}{\bf
W}^H{\bf C}_{\tau_k}{\bf X}E\{{\bf V}_{k}\}{\bf X}^H{\bf
C}^H_{\tau_k}{\bf W}{{\bf \Phi }}^H\}
%\nonumber\\&\approx \frac{ |\beta_k|^2M}{\tilde{M}}\mathrm{Tr}\{{\bf
%W}^H{\bf
%C}_{\tau_k}{\bf X}{\bf X}^H{\bf C}^H_{\tau_k}{\bf W }\}.
\end{align}
where ${\bf V}_{k}={\bf v}(\theta_k){\bf v}^H(\theta_k)$ and its
$(i,j)$-th entry  can be expressed as $V_k(i,j)=e^{j\frac{2\pi
f}{c}(r^t_i\cos(\theta_k-\alpha_i)-r^t_j\cos(\theta_k-\alpha_j))}$.
As already noted, the Doppler shift within a pulse is ignored in
(\ref{sig_power_k}).
 Since the nodes are uniformly dispersed on
a disk of radius $r$,
%the probability density functions of the $i$-th TX/RX node location
%$(r^{t/r}_i,\alpha^{t/r}_i)$ \cite{Ochiai} are
%\begin{eqnarray}
%f_{r_i^{t/r}}(r_i^{t/r})&=&\frac{2r_i^{t/r}}{r^2}, \
%0<r_i^{t/r}<r\nonumber \\
%\mathrm{and} \ f_{\alpha_i^{t/r}}(\alpha_i^{t/r})&=&\frac{1}{2\pi},\
%-\pi\leq\alpha_i^{t/r}<\pi. \label{2}
%\end{eqnarray}
%Then
the distribution of
$h\triangleq\frac{r_i^{t/r}}{r}\sin(\alpha_i^{t/r}-\psi_0)$ is given
by \cite{Ochiai}
\begin{eqnarray}
f_h(h)=\frac{2}{\pi}\sqrt{1-h^2},-1<h<1
\end{eqnarray}
so that
\begin{eqnarray}\label{property}
E\left\{e^{j\alpha h}\right\}=2\frac{J_1(\alpha)}{\alpha}
\end{eqnarray}
 where
$J_1(\cdot)$ is the first-order Bessel function of the first kind.
 Thus, based on (\ref{property}) we can obtain
\cite{Yu:09_tsp}
 \begin{eqnarray}
&&E\left\{e^{j\frac{2\pi
f}{c}(r^t_i\cos(\theta_k-\alpha_i)-r^t_j\cos(\theta_{k'}-\alpha_j))}\right\}
\nonumber\\&=& \left\{
\begin{array}{rl}
1& i=j\ \text{and}\ k=k'\\
\varsigma(4\sin(\frac{\theta_{k'}-\theta_k}{2}))& i=j\ \text{and}\
k\neq
k'\\
\varsigma^2(2) &  i\neq j
\end{array} \right.
 \end{eqnarray}
where $\varsigma(x)=2\frac{J_1(x\frac{\pi rf}{c})}{x\frac{\pi
rf}{c}}$.  As observed in \cite{Yu:09_tsp}, the terms multiplied by
$\varsigma^2(2)$ are small enough and can be neglected. Thus, the
average power $ P^k_s$ in (\ref{sig_power_k}) can be further
approximated by
\begin{align}\label{sig_power_k1}
 P^k_s&\approx
|\beta_k|^2N_pN_r\mathrm{Tr}\{{\bf \Phi}{\bf W}^H{\bf
C}_{\tau_k}{\bf X}{\bf X}^H{\bf C}^H_{\tau_k}{\bf
W}{\bf \Phi }^H\}\nonumber\\
&\approx \frac{ |\beta_k|^2MN_pN_r}{\tilde{M}}\mathrm{Tr}\{{\bf
W}^H{\bf C}_{\tau_k}{\bf X}{\bf X}^H{\bf C}^H_{\tau_k}{\bf W }\}.
\end{align}

Inserting ${\bf \Phi}_{\#2}$ into (\ref{inter_power}),  the average
power of the interference  can be approximated as
\begin{align}\label{inter_power1}
 P_n&=\sigma^2N_pN_r\mathrm{Tr}\{{\bf \Phi W}^H{\bf W\Phi
 }^H\}
 =\sigma^2N_pN_r\sum_{q=1}^{M}\sum_{i,j}{\Phi}_{qi}w_{ij}{ \Phi}^*_{qj}
 \approx
 \sigma^2N_pN_rM
\end{align}
where ${\Phi}_{ij}$ and $w_{ij}$ are the $(i,j)$-th entries of ${\bf
\Phi}$ and ${\bf  W}^H{\bf W
 }$, respectively. The approximation in  (\ref{inter_power1})
uses the constraint  $\mathrm{diag}\{{\bf W}^H{\bf
W}\}=[1,\ldots,1]^T$ and the fact that $\sum_{q=1}^{M}\sum_{i\neq
j}{\Phi}_{qi}w_{ij}{ \Phi}^*_{qj}\approx 0$ for sufficiently large
$\tilde{M}$ due to ${\Phi}_{qi}\sim \mathcal{N}(0,1/\tilde{M})$.

Based on (\ref{sig_power_k1}) and (\ref{inter_power1}), the SIR is
given approximately by
\begin{align}\label{SIR}
SIR_k=P^k_s/ P_n
 \approx
 \frac{ |\beta_k|^2}{\sigma^2\tilde{M}}\mathrm{Tr}\{{\bf  W}^H{\bf
Q}_{\tau_k}{\bf W }\}
\end{align}
where ${\bf Q}_{\tau_k}={\bf C}_{\tau_k}{\bf X}{\bf X}^H{\bf
C}^H_{\tau_k}$ is an $(L+\tilde{L})\times (L+\tilde{L})$ matrix of
rank $M_t$. The
 maximization of $SIR_k$  over ${\bf W}$ can thus be approximated
 by the problem
\begin{align}\label{SIR_pro}
{\bf W}^*&=\max_{{\bf W},\tilde{M}}
 \frac{ |\beta_k|^2}{\sigma^2\tilde{M}}\mathrm{Tr}\{{\bf  W}^H{\bf
Q}_{\tau_k}{\bf W
 }\}\nonumber\\
 &s.t.\ \mathrm{diag}\{{\bf W}^H{\bf W}\}=[1,\ldots,1]^T_{\tilde{M}\times
1}.
\end{align}

It can be easily seen that  ${\bf W}^*$ contains as its columns the
eigenvectors corresponding to the largest eigenvalue of $ {\bf
Q}_{\tau_k}$. Since the largest eigenvalue of ${\bf Q}_{\tau_k}$ is
not greater than $\mathrm{Tr}\{{\bf Q}_{\tau_k}\}=M_t$, the maximum
$SIR_k$ is bounded by
\begin{align}\label{SIR_bound}
Bound\ 1:\  \frac{ |\beta_k|^2}{\sigma^2}\leq SIR_k\leq \frac{
 |\beta_k|^2M_t}{\sigma^2}.
\end{align}
The upper bound is achieved when the rank of $\bf X$  equals  $1$,
i.e., all the transmit nodes send out the same waveforms. When
orthogonal waveforms are utilized, i.e., ${\bf X}^H{\bf X}={\bf
I}_{M_t}$, the $SIR_k$ reaches the lower bound.

%However, the solution ${\bf W}^*$ would invalidate the conditions
%for the application of CS since a pulse is equivalently compressed
%to a single measurement. Fortunately, there are multiple equally
%large eigenvalues

It can be shown that, when the transmit waveform are orthogonal, i.e.,  ${\bf X}^H{\bf X}={\bf I}_{Mt}$,
  ${\bf Q}_{\tau_k}$ has $M_t$ nonzero eigenvalues which are all
equal to $1$. Therefore, for a fixed $\tilde{M}$, $\tilde{M}\leq
M_t$, the optimal ${\bf W}$ contains the $\tilde{M}$ eigenvectors of
${\bf Q_k}$ corresponding to eigenvalue $1$ and achieves maximum
$SIR_k$ equal to $SIR_k=\frac{ |\beta_k|^2}{\sigma^2}$. Since the
maximum $SIR_k$ is independent of $\tilde{M}$, any matrix containing
$\tilde{M}, \tilde{M}\leq M_t$, eigenvectors of ${\bf Q}$
corresponding to eigenvalue $1$ would give rise to the maximum
$SIR_k$. However, $\tilde{M}=M_t$ results in smaller CSM than any
$\tilde{M}$ less than $M_t$ due to the fact that  the rank of  ${\bf
W}$ is $\tilde{M}$. Therefore, the optimal ${\bf W}$ is
\begin{align}\label{opti_SIR}
{\bf W}^{**}={\bf C}_{\tau_k}{\bf X}.
\end{align}
For the case of completely coherent transmit waveforms in which the
upper bound in (\ref{SIR_bound}) is achieved,  the resulting ${\bf
W}^{**}$ is rank deficient.

Unfortunately, ${\bf W}^{**}$ is not achievable  since it depends on
the time delay induced by a target, $\tau_k$, which  is unknown. To
address this issue, we replace $SIR_k$ in  the objective function in
(\ref{SIR_pro}) with the average $SIR_k$, where the average is taken
over all possible delays, and is denoted here by $\overline{SIR_k}$.
Assuming that the time delay induced by the $k$-th target follows a
discrete uniform distribution, i.e.,
$p(\tau_k=k)=\frac{1}{\tilde{L}+1},k=0, \ldots,\tilde{L}$, we can
write
\begin{align}\label{SIR_new}
\overline{SIR}_k&=
 \frac{
|\beta_k|^2}{\sigma^2\tilde{M}}\sum_{\tau=0}^{\tilde{L}}\frac{1}{\tilde{L}+1
}\mathrm{Tr}\{{\bf  W}^H{\bf Q}_{\tau}{\bf W
 }\}=\frac{
|\beta_k|^2}{\sigma^2\tilde{M}}\frac{1}{\tilde{L}+1}\mathrm{Tr}\{{\bf
W}^H{\bf C}{\bf W
 }\}
\end{align}
where
\begin{align}
{\bf C}=\sum_{\tau=0}^{\tilde{L}}{\bf Q}_{\tau}=[{\bf C}_{0}{\bf
X},\ldots,{\bf C}_{\tilde{L}}{\bf X}][{\bf C}_{0}{\bf X},\ldots,{\bf
C}_{\tilde{L}}{\bf X}]^H.
\end{align}
Therefore, the optimization problem that maximizes
$\overline{SIR}_k$ can be rewritten as
\begin{align}\label{SIR_pro1}
{\bf W}^*&=\max_{{\bf W},\tilde{M}}\overline{SIR}_k
 \nonumber\\
 &s.t.\ \mathrm{diag}\{{\bf W}^H{\bf W}\}=[1,\ldots,1]^T_{\tilde{M}\times
1}.
\end{align}

The solution ${\bf W}^*$ of the above problem  contains as its
columns the eigenvectors corresponding to the largest eigenvalue of
$ {\bf C}$. Unlike (\ref{SIR_pro}), we cannot find a close-form
solution to (\ref{SIR_pro1}) that has sufficiently high rank.
Further, the problem of (\ref{SIR_pro1}) is non-convex. Inspired by
the form of (\ref{opti_SIR}), we propose a feasible  ${\bf W}$ by
taking all possible delays into account as follows:
\begin{align}\label{subopt_2}
{\bf W}=[{\bf C}_{0}{\bf X},\ldots, {\bf C}_{\tilde{L}}{\bf X} ].
\end{align}
Since ${\bf C}_{i}{\bf X}$ contains eigenvectors  corresponding  to
the largest eigenvalues of ${\bf Q}_{i}$,  utilizing
(\ref{subopt_2}) results in the average $SIR_k$ bounded by
\begin{align}\label{SIR_bound1}
Bound\ 2:\ \frac{
|\beta_k|^2}{\sigma^2}\frac{1}{\tilde{L}+1}+\Delta\leq\overline{SIR}_k\leq
\frac{ |\beta_k|^2}{\sigma^2}\frac{M_t}{\tilde{L}+1}+\Delta
\end{align}
where $\Delta$  denotes
$\frac{|\beta_k|^2}{\sigma^2M_t(\tilde{L}+1)^2}\mathrm{Tr}\{\sum_{\tau'\neq
\tau}{\bf X}^H{\bf C}^H_{\tau}{\bf Q}_{\tau'}{\bf C}_{\tau}{\bf
X}\}$. One can see that \emph{Bound} 2 would be reduced to
\emph{Bound} 1 when $\tilde{L}=0$.

%For a fixed $\tilde{L}$, \emph{Bound} 2  would be pushed up to its
%limit \emph{Bound} 1 if  ${\bf Q}_{0}=\ldots={\bf
%Q}_{\tau}=\ldots={\bf Q}_{\tilde{L}}$. Recall that ${\bf Q}_{\tau}$
%is a square matrix of size $L+\tilde{L}$, formed by shifting ${\bf
%X}{\bf X}^H$ down along its diagonal by $\tau$. Therefore, it is
%impossible for ${\bf Q}_{\tau},\ \tau=0,\ldots,\tilde{L}$, to
%satisfy the aforementioned condition. However,   the waveforms $\bf
%X$ that results in ${\bf X}{\bf X}^H$ with a large number of
%identical entries would significantly improve \emph{Bound} 2. On can
%easily see that  ${\bf X}{\bf X}^H$ would have the largest number of
%identical entries if all the nodes transmit the same rectangular
%pulse and thus the maximum $\overline{SIR}_k$ can be achieved. On
%the contrary,  the worst $\overline{SIR}_k$ will be achieved by
%using the independent and orthogonal waveforms, e.g., independently
%generated quadrature phase shift
% keying (QPSK) waveforms,  though not strictly orthogonal.

Next, we  examine  the resulting SIR based on three types of
waveforms, namely a rectangular pulse, independently generated
quadrature phase shift
 keying (QPSK) waveforms
and Hadamard codes.
In particular, we show that  using ${\bf \Phi}_{\#2}$ can
suppress interference uncorrelated with the transmit waveforms, and
maintains coherence as low as that corresponding to the Gaussian random measurement matrix.

\subsubsection{SIR under the conventional measurement matrix} Let us
consider a conventional measurement matrix ${\bf \Phi}_c$, which is
an $M\times (L+\tilde{L})$ Gaussian random matrix of unit column
norm with $\mathrm{Tr}\{{\bf \Phi}_c{\bf \Phi}_c^H\}=M$.
 The average power of
the interference is $P_n= \sigma^2M$  (see  (\ref{inter_power})).

 Let
$\mathbf{S}_i$ be a square matrix, formed by shifting the main
diagonal of $\mathbf{I}_L$ up by $i$ positions. It can be easily
seen that $\mathbf{S}^H_i=\mathbf{S}_{-i}$.  The average power of
the target returns from $K$ targets at a receive node, conditioned
on the transmit waveforms, equals
\begin{align}\label{sig_power}
{P}_s&=E\{{\bf r}^H{\bf r} | {\bf X}\}=\sum_k
 P_s^k+\sum_{k\neq k'}P_s^{k,k'}
\end{align}
where
\begin{align}
P_s^k
 &= |\beta_k|^2E\{\mathrm{Tr}\{{{\bf \Phi }}_c{\bf
C}_{\tau_k}{\bf X}{\bf v}(\theta_k)({{\bf \Phi }}{\bf
C}_{\tau_k}{\bf X}{\bf v}(\theta_k))^H\}\}\nonumber\\
&\approx |\beta_k|^2\mathrm{Tr}\{{\bf \Phi}_c{\bf C}_{\tau_k}{\bf
X}{\bf X}^H{\bf C}^H_{\tau_k}{\bf \Phi
}_c^H\}\nonumber\\
&\approx \frac{M_tM|\beta_k|^2}{L+\tilde{L}}\\
\mathrm{and}\nonumber\\
 P_s^{k,k'}
&\approx
\underbrace{\beta_k^*\beta_{k'}\varsigma^2\left(4\sin\left(\frac{\theta_{k}-
\theta_{k'}}{2}\right)\right)e^{\frac{4\pi
f(d_k(0)-d_{k'}(0))}{c}}}_{\gamma_{kk'}}\frac{M}{L+\tilde{L}}\mathrm{Tr}\{{
\bf X}^H{\bf S}_{\tau_k-\tau_{k'}}{\bf X}\}.
\end{align}

\subsubsection{SIR for the  measurement matrix ${\bf \Phi}_{\#2}$}
The proposed measurement matrix ${{\bf \Phi}}_{\#2}={\bf \Phi W}^H$
results in the same average interference power
 as the matrix ${{\bf
\Phi}}_c$. The average power of the desired signal conditioned on
the transmit waveforms, $\tilde P_s$, however, will improve. Like
(\ref{sig_power}), $\tilde{P}_s$ can be partitioned into the sum of
the autocorrelation, $\tilde{P}_s^k$, and  cross correlation,
$\tilde{P}_s^{k,k'}$, of the  returns from $K$ targets. It holds
that
\begin{align}\label{sig_power_one}
\tilde{P}_s^k
 &= |\beta_k|^2E\{\mathrm{Tr}\{\tilde{{\bf \Phi }}_{\#2}{\bf
C}_{\tau_k}{\bf X}{\bf v}(\theta_k)(\tilde{{\bf \Phi }}_{\#2}{\bf
C}_{\tau_k}{\bf X}{\bf v}(\theta_k))^H\}\}
\nonumber\\
 &\approx  |\beta_k|^2\mathrm{Tr}\{{\bf \Phi }{\bf W}^H{\bf
C}_{\tau_k}{\bf X}{\bf X}^H{\bf C}^H_{\tau_k}{\bf
W}{\bf \Phi}^H\}\nonumber\\
&\approx\frac{|\beta_k|^2M}{(\tilde{L}+1)M_t}\mathrm{Tr}\{{\bf
W}^H{\bf C}_{\tau_k}{\bf X}{\bf X}^H{\bf C}^H_{\tau_k}{\bf
W}\}\nonumber\\
&=\frac{|\beta_k|^2M}{(\tilde{L}+1)M_t}
\sum_{q=0}^{\tilde{L}}\mathrm{Tr}\{{\bf X}^H{\bf S}_{q-\tau_k}{\bf
X}{\bf
X}^H{\bf S}^H_{q-\tau_k}{\bf X}\}\\
\mathrm{and}\nonumber\\
 P_s^{k,k'}
&\approx
\frac{\gamma_{kk'}M}{(\tilde{L}+1)M_t}\sum_{q=0}^{\tilde{L}}\mathrm{Tr}\{{\bf
X}^H{\bf S}_{\tau_k-q}{\bf X}{\bf X}^H{\bf S}_{q-\tau_{k'}}{\bf
X}\}.
\end{align}

For  orthogonal, or randomly generated waveforms across the transmit
nodes, $\tilde{P}_s^k$ always dominates  the average power of the
desired signal.
 In order to increase $\tilde{P}^k_s$, the quantity $\mathrm{Tr}\{{\bf
X}^H{\bf S}_{q-\tau_k}{\bf X}{\bf X}^H{\bf S}^H_{q-\tau_k}{\bf X}\}$
in (\ref{sig_power_one}) needs to be as large as possible. ${\bf
X}^H{\bf S}_{m}{\bf X}$ can be  expressed as
\begin{align}\label{seq_cor}
{\bf X}^H{\bf S}_{m}{\bf X}= \left\{
\begin{array}{rl}
{\bf X}^H_{1:L-m}{\bf X}_{m+1:L},  &  m\geq0 \\
{\bf X}^H_{1-m:L}{\bf X}_{1:L+m},  & \text{otherwise}
\end{array} \right.
\end{align}
where ${\bf X}_{i:j}$ denotes the matrix that contains the rows of
${\bf X}$ indexed from $i$ to $j$.

Eq. (\ref{seq_cor}) implies that the  non-circular autocorrelation
of the waveform sequence of a transmit node, i.e.,
$R_i(\tau)=\int_{t=0}^{T_p}x_i(t)x_i^*(t-\tau), i=1,\ldots,M_t$,
should be insensitive to the shift.  This essentially  requires a
narrowband signal.     Based on this principle, the best candidate
is a rectangular pulse and the maximum $\tilde{P}_s$  equals
\begin{align}\label{sinr_rec}
\tilde{P}_s &=\frac{MM_t}{(\tilde{L}+1)}
\sum_{k=1}^K\sum_{q=0}^{\tilde{L}}|\beta_k|^2\left(\frac{L-|q-\tau_k|}{L}
\right)^2+ \frac{MM_t}{(\tilde{L}+1)}\sum_{k\neq
k'}\sum_{q=0}^{\tilde{L}}\gamma_{kk'}\frac{(L-|\tau_k-q|)(L-|q-\tau_{k'}|)}{
L^2}\nonumber\\
&\leq MM_t \left(\sum_{k=1}^K|\beta_k|^2+ \sum_{k\neq
k'}\gamma_{kk'}\right).
\end{align}
 The equality in (\ref{sinr_rec}) holds only if the targets induce  identical
delays and the sampling window  is of a length that exactly covers
the duration of target returns.
 Obviously,  the transmit nodes cannot use
identical waveforms. This is because  the transmit waveforms are
required to be orthogonal, or randomly generated in order to
maintain low CSM.

Similarly, the minimum  average power of the desired signal is
achieved when  randomly generated QPSK waveforms are used, because
such waveforms cover the widest bandwidth for the fixed pulsed
duration $T_p$ and the length of waveforms $L$. The corresponding
value of  $\tilde{P}_s$ is approximately
\begin{align}\label{sinr_random}
\tilde{P}_s & \approx \frac{M}{(\tilde{L}+1)}
\sum_{k=1}^K|\beta_k|^2\left(\sum_{q=0,q\neq
\tau_k}^{\tilde{L}}M_t\frac{L-|q-\tau_k|}{L^2}+1\right).
\end{align}
%One can see that (\ref{sinr_rec}) and (\ref{sinr_random}) coincide
%with our aforementioned intuitive observations.

For  orthogonal Hadamard waveforms that are of smaller bandwidth
than the randomly generated QPSK waveforms,  the average power of
the desired signal equals approximately
\begin{align}\label{sinr_hadamard}
\tilde{P}_s & \approx \frac{M}{(\tilde{L}+1)}
\sum_{k=1}^K\sum_{q=0}^{\tilde{L}}|\beta_k|^2\left(\frac{L-|q-\tau_k|}{L}\right)^2.
\end{align}
Recall that ${\bf W}^*$ corresponding to the true delay gives rise
to the maximum received signal power. Adding the terms ${\bf
C_{\tilde{\tau}_k}}{\bf X}, \tilde{\tau}_k\neq \tau_k$ to ${\bf W}$
(see (\ref{subopt_2})) would
 lower $\tilde{P}_s^k$. When a coarse delay estimate is available, we  need
to  consider  only delays around that estimate and thus the length
of the sampling window can be shortened. This effectively reduces
the number of possible delays that are
 considered for the construction of ${\bf W}$.   Therefore,
$\tilde{P}_s^k$ can be improved
 for the
waveforms considered above if a coarse delay estimate is available.

\subsubsection{The SIR gain}

Let $SIR_p$ and $SIR_c$ denote the SIR
corresponding to measurement matrices  ${\bf \Phi}_{\#2}$  and ${\bf \Phi}_{c}$, respectively. When transmitting Hadamard codes, the SIR
gain induced by using the proposed measurement matrix can be
expressed as
\begin{align}
r_{Ha}&=\frac{P_s}{\tilde{P}_s} \approx
\frac{\frac{M}{(\tilde{L}+1)}
\sum_{k=1}^K\sum_{q=0}^{\tilde{L}}|\beta_k|^2\left(\frac{L-|q-\tau_k|}{L}\right)^2}{\sum_{k=1}^K\frac{M_tM|\beta_k|^2}{L+\tilde{L}}}\nonumber\\
&=\frac{L+\tilde{L}}{M_t(\tilde{L}+1)L^2}\frac{
\sum_{k=1}^K|\beta_k|^2C_{k}}{\sum_{k=1}^K|\beta_k|^2}
\end{align}
where
\begin{align}
C_{k}&=\sum_{q=0}^{\tilde{L}}\left(L-|q-\tau_k|\right)^2=(\tilde{L}+1-2L)(\tau_k-\tilde{L}/2)^2
+\sum_{q=L-\tilde{L}}^{L}q^2+\frac{(2L-\tilde{L}-1)\tilde{L}^2}{4}.
\end{align}
For a fixed $\tilde{L}$ and $L$, with $0\leq\tilde{L}\leq2L-1$,
$C_{k}$ can be bounded as
\begin{align}
\sum_{q=L-\tilde{L}}^{L}q^2\leq C_{k}\leq
\sum_{q=L-\tilde{L}}^{L}q^2+\frac{(2L-\tilde{L}-1)\tilde{L}^2}{4}.
\end{align}
Therefore,   lower and upper bounds on the approximate SIR gain
using Hadamard codes are given by
\begin{align}
\frac{(L+\tilde{L})\sum_{q=L-\tilde{L}}^{L}q^2}{M_t(\tilde{L}+1)L^2}
\leq
r_{Ha}\leq\frac{(L+\tilde{L})(\sum_{q=L-\tilde{L}}^{L}q^2+\frac{(2L-\tilde{L
}-1)\tilde{L}^2}{4})}{M_t(\tilde{L}+1)L^2}.
\end{align}
% It can be seen from the expression above that the SIR gain is
%maximized when the sampling window  of length $L$  exactly covers
%the duration of target returns. In addition to this condition, it
%requires  targets to induce  identical delay.
%For the slowly
%moving targets, the conclusions of SIR still approximately hold.
Similarly,  the SIR gain using  randomly generated QPSK waveforms is
bounded by
\begin{align}\frac{(L+\tilde{L})(\sum_{q=L-\tilde{L}}^{L}q+\frac{L^2}{M_t}-L
) }{(\tilde{L}+1)L^2}\leq r_{QPSK}&\leq
\frac{(L+\tilde{L})(\sum_{q=L-\tilde{L}}^{L}q+\frac{\tilde{L}^2}{4}+\frac{L^
2}{M_t}-L)}{(\tilde{L}+1)L^2}.
\end{align}
As long as $\tilde{L}<L$ and $M_t<L$, $r_{QPSK}$  is always greater
than 1. When
$\frac{\sum_{q=L-\tilde{L}}^{L}q^2}{M_t\sum_{q=L-\tilde{L}}^{L}q}>1$,
the lower bound on $r_{Ha}$ is higher than that on  $r_{QPSK}$. For
a sufficiently long $L$ and moderate $M_t$,  $r_{Ha}$ would be
superior to $r_{QPSK}$.  Based on (\ref{sinr_rec}) and
(\ref{sinr_hadamard}), one can infer that the SIR gain using the
rectangular pulse is approximately $M_t$ times greater than that
using Hadamard codes.
\subsubsection{The CSM based
on the suboptimal measurement matrix \#2 }\label{coherence_sopt2}

%As stated in the beginning of this section,  the measurement matrix
%${\bf \Phi}_{\#2}$ with a proper ${\bf W}$ may  not increase the
%coherence of the sensing matrix as compared to the Gaussian random
%measurement matrix.
In this section, we  examine the effect of the proposed ${\bf W}$ in
(\ref{subopt_2}) on the CSM. For simplicity, the targets are
considered to be stationary and  the possible delays for
constructing ${\bf W}$ are based on the range grid points used to
form the basis matrix. Then the sensing matrix based on ${\bf
\Phi}_{\#2}$, or the Gaussian random matrix can be respectively
represented as
\begin{align}
{{\bf \Theta}}={\bf \Phi}_{\#2}\mathbf{\Psi}={\bf
\Phi}_{M\times(\tilde{L}+1)M_t}{\bf W}^H{\bf W}{\bf V}
\end{align}
and
\begin{align}
\tilde{{{\bf \Theta}}}={\bf
\Phi}_{M\times(L+\tilde{L})}\mathbf{\Psi}={\bf
\Phi}_{M\times(L+\tilde{L})}{\bf W}{\bf V}
\end{align}
 where ${\bf V}={kron}({\bf
I}_{\tilde{L}+1},[{\bf v}(a_1),\ldots,{\bf v}(a_{N_a})])$ and ${\bf
\Phi}_{i\times j}$ is an $i\times j$ Gaussian random matrix whose
entries are of zero mean and variance $1/j$. For sufficiently large
$j$, the column coherence of ${{\bf \Theta}}$  can be approximated
as
\begin{align}\label{coherence}
\mu_{kk'}({{\bf \Theta}})=\frac{\left|\sum_i(\sum_m
\Phi(m,i)\Phi^*(m,i))v_{kk'}(i)\right|}{\sqrt{\sum_i(\sum_m
\Phi(m,i)\Phi^*(m,i))v_{kk}(i)\sum_i(\sum_m
\Phi(m,i)\Phi^*(m,i))v_{k'k'}(i)}}
\end{align}
where $v_{kk'}(i)$ denotes the $i$-th diagonal element of the matrix
${\bf W}^H{\bf C}_{\lfloor \frac{2c_{k}}{cT_p/L}\rfloor}{\bf X}{\bf
v}_m(a_{k})\left({\bf W}^H{\bf C}_{\lfloor
\frac{2c_{k'}}{cT_p/L}\rfloor}{\bf X}{\bf v}_m(a_{k'})\right)^H$.
Without loss of generality, we let the columns of $\bf \Phi$ be of
unit norm. Then (\ref{coherence}) can be further written as
\begin{align}\label{coherence_1}
\mu_{kk'}({{\bf \Theta}})=\frac{\left|\sum_i
v_{kk'}(i)\right|}{\sqrt{\sum_i v_{kk}(i)\sum_i v_{k'k'}(i)}}.
\end{align}
One can easily see from (\ref{coherence_1}) that the coherence of
${{\bf \Theta}}$ is approximately equal to that of  matrix ${\bf
W}^H{\bf W}{\bf V}$. The same conclusion applies to $\tilde{\bf
\Theta}$ as well, i.e., the coherence of $\tilde{{\bf \Theta}}$ is
approximately equal to that of  matrix ${\bf W}{\bf V}$. Since ${\bf
W}^H{\bf W}$ is more ill-conditioned than ${\bf W}$, the conditional
number of ${\bf W}^H{\bf W}{\bf V}$ is greater than that of ${\bf
W}^H{\bf W}$. Therefore, using ${\bf \Phi}_{\#2}$ increases the
maximum CSM as compared to the Gaussian random measurement matrix
with high probability. However, for a well conditioned ${\bf W}$,
the increase of the maximum CSM caused by ${\bf \Phi}_{\#2}$ is
negligible.

\subsection{$\mathbf{\Phi}_{\#1}$ v.s. $\mathbf{\Phi}_{\#2}$}
We have proposed two measurement matrices based on different
performance metrics. The advantages and disadvantages of
$\mathbf{\Phi}_{\#1}$ and ${\bf \Phi}_{\#2}$ are summarized as
follows.
\begin{itemize}
\item \emph{Complexity}

Solving $\mathbf{\Phi}_{\#1}$ involves  a complex optimization
problem and depends on a particular basis matrix, while
$\mathbf{\Phi}_{\#2}$  requires knowledge only of all the possible
discretized time delays. Therefore, the construction of
$\mathbf{\Phi}_{\#1}$ involves  higher computational complexity than
does $\mathbf{\Phi}_{\#2}$.

\item \emph{Performance}

$\mathbf{\Phi}_{\#1}$ aims at decreasing the coherence of the
sensing matrix and enhancing SIR simultaneously.
%$\Phi_{\#2}$ targets
%SIR improvement only and keep the coherence of the sensing matrix as
%low as the Gaussian random measurement matrix.
The tradeoff between CSM and SIR   results in $\mathbf{\Phi}_{\#1}$
yielding lower SIR than  $\mathbf{\Phi}_{\#2}$.  Therefore,
$\mathbf{\Phi}_{\#1}$  is expected to perform better than
$\mathbf{\Phi}_{\#2}$ in the case of low interference,
  while it
should  perform worse in the presence of strong interference.
\end{itemize}

%%%%%%%%%%%%%%%%%%%%%%%%%%%%%%%%%%%%%%%%%%%%%%%%%%%%%%%%%%%%%%%%%%%%%%%%%%%%
%%%%%%%%%%%%%%%%%%%%%%%%%%%%%%%%%%%%%%%%%%%%%%%%%%%%%%%%%%%%%%
\section{Simulation Results}\label{simulation}
In this section, we demonstrate the performance of CS-based MIMO radar when
using the proposed measurement matrices ${\bf \Phi}_{\#1}$ and ${\bf
\Phi}_{\#2}$, respectively. We consider a MIMO radar system with
transmit and receive nodes uniformly located on a disk of radius
$10$m.   The carrier frequency is $f=5 GHz$.
%Each transmit node uses %orthogonal quadrature phase shift keying (QPSK)
%a waveform sequence of length $L=512$ and unit power.
The received signal is
 corrupted by  zero-mean Gaussian noise. The
signal-to-noise ratio (SNR)  is defined as the inverse  of the power
of thermal noise at a receive node. A jammer is located at angle $7
\textordmasculine$ and transmits an unknown Gaussian random
waveform. The targets are assumed to fall on the grid points.

\subsection{The proposed measurement matrix ${\bf \Phi}_{\#2}$}
\subsubsection{SIR improvement}
\begin{figure}[htbp]
\centerline{\epsfig{figure=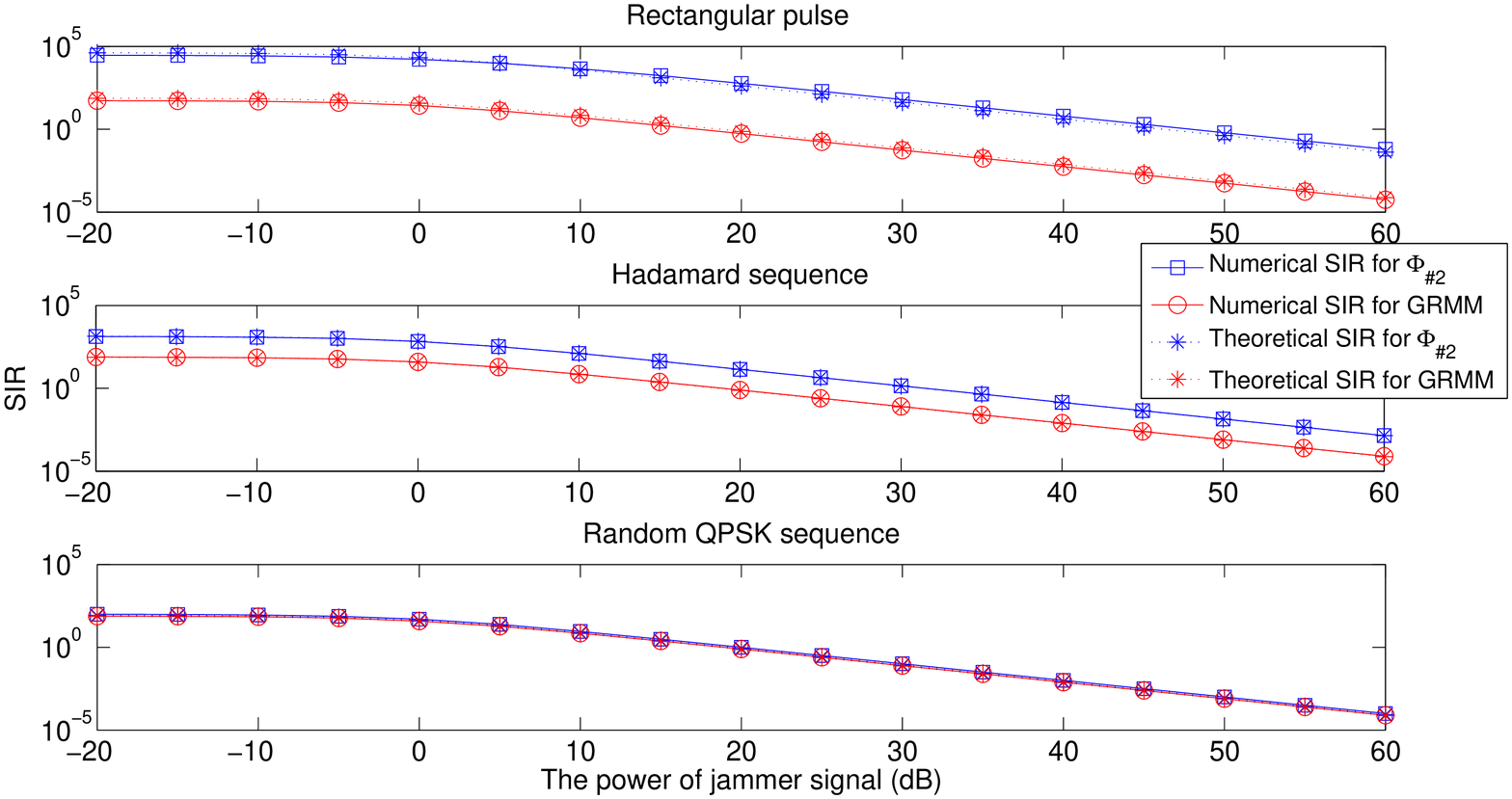,height=3.5in,width=12.0cm}
} \caption{{SIR corresponding to GRMM and $\bf{\Phi}_{\#2}$ for
different transmit waveforms ($M=M_t=30$ and $N_r=1$).
 }}
 \label{SINR}
\end{figure}

\begin{figure}[htbp]
 \centerline{\epsfig{figure=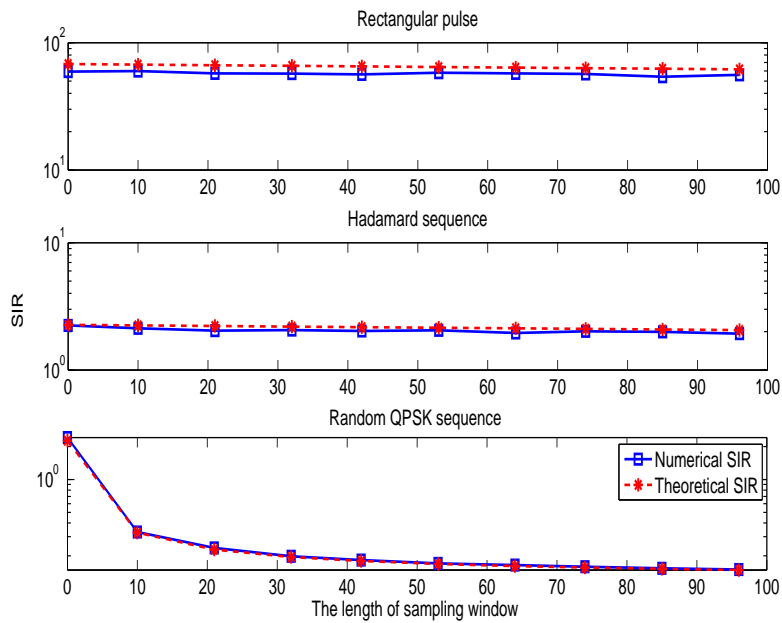,height=3.5in,width=12.0cm}}
\caption{{SIR corresponding to   ${\bf \Phi}_{\#2}$
 for different values of $\tilde{L}$ ($M=M_t=30$ and $N_r=1$).
 }}
 \label{SIR_L1}
\end{figure}

$M=30$ compressed measurements are forwarded to the fusion center by
each receive node.  The maximum possible delay  is $\tilde{L}=100$.
 Figure \ref{SINR} compares  the numerical and theoretical  SIR
 produced using the rectangular-pulse,  Hadamard  waveforms and randomly
generated  QPSK
 waveforms for the case of $M_t=30$ transmit nodes and $N_r=1$ receive node.
The SIR performance, shown in Fig. \ref{SINR}, is the average
 of  $1000$ independent and random runs.
 The theoretical SIR of these three  sequences is calculated  based on
(\ref{sinr_rec}), (\ref{sinr_hadamard}) and
 (\ref{sinr_random}), respectively. The power of thermal noise is fixed to $1$ and
 the power of the jammer varies from $-20$dB to $60$dB.
 %It can seen from
% Fig. {\ref{SINR}} that the various  sequences will yield the same SINR
%using the conventional  measurement matrix ${\bf \Phi}_c$.
%This is because the conventional measurement matrix is a Gaussian
%random matrix that is independent of the transmit waveforms.
Applying the proposed measurement ${\bf \Phi}_{\#2}$ at the
receivers, the
 rectangular pulse and Hadamard waveforms produce a significant  SIR gain
over the Gaussian random measurement matrix (GRMM), while the random
QPSK
 sequence achieves almost no gain. Furthermore, the numerical SIR
performance follows the theoretical SIR   for all three sequences.
Figure \ref{SIR_L1} demonstrates the SIR performance obtained by averaging over
 $500$ independent  runs corresponding to independent interference waveforms,
 for different values of the
 maximum time delay
 $\tilde{L}$.  We consider a case in which  only one target exists   and the
jammer power is 225.
 One can see that  a decrease in $\tilde{L}$ can
 significantly improve SIR yielded by QPSK waveforms when $\tilde{L}$ is
less than $10$.  In contrast,
 Hadamard waveforms and rectangular pulse produce almost the same SIR
 for different values of
 $\tilde{L}$. This indicates that for QPSK waveforms the prior information of possible
 delays enables SIR improvement, while for the other two types of waveforms considered prior information did not make a difference.

\subsubsection{The CSM}
\begin{figure}[htbp]
\centerline{\epsfig{figure=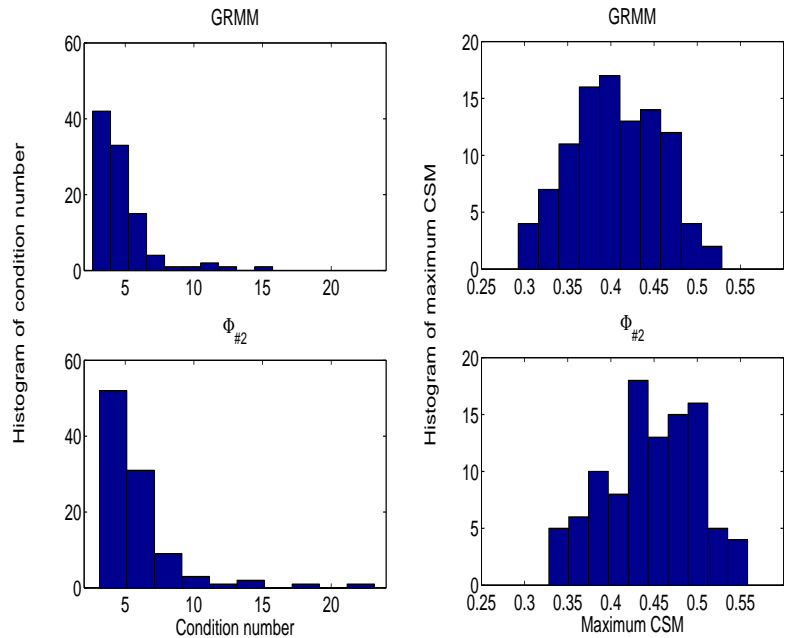,height=3.5in,width=12.0cm}}
\caption{{ Conditional number and the maximum coherence of the
sensing
 matrix based on ${\bf \Phi}_{\#2}$
 ($M=30$ and $N_r=M_t=10$).
 }}
 \label{ConditionaNumber_MaximumCoherence}
\end{figure}

\begin{figure}[htbp]
\centerline{\epsfig{figure=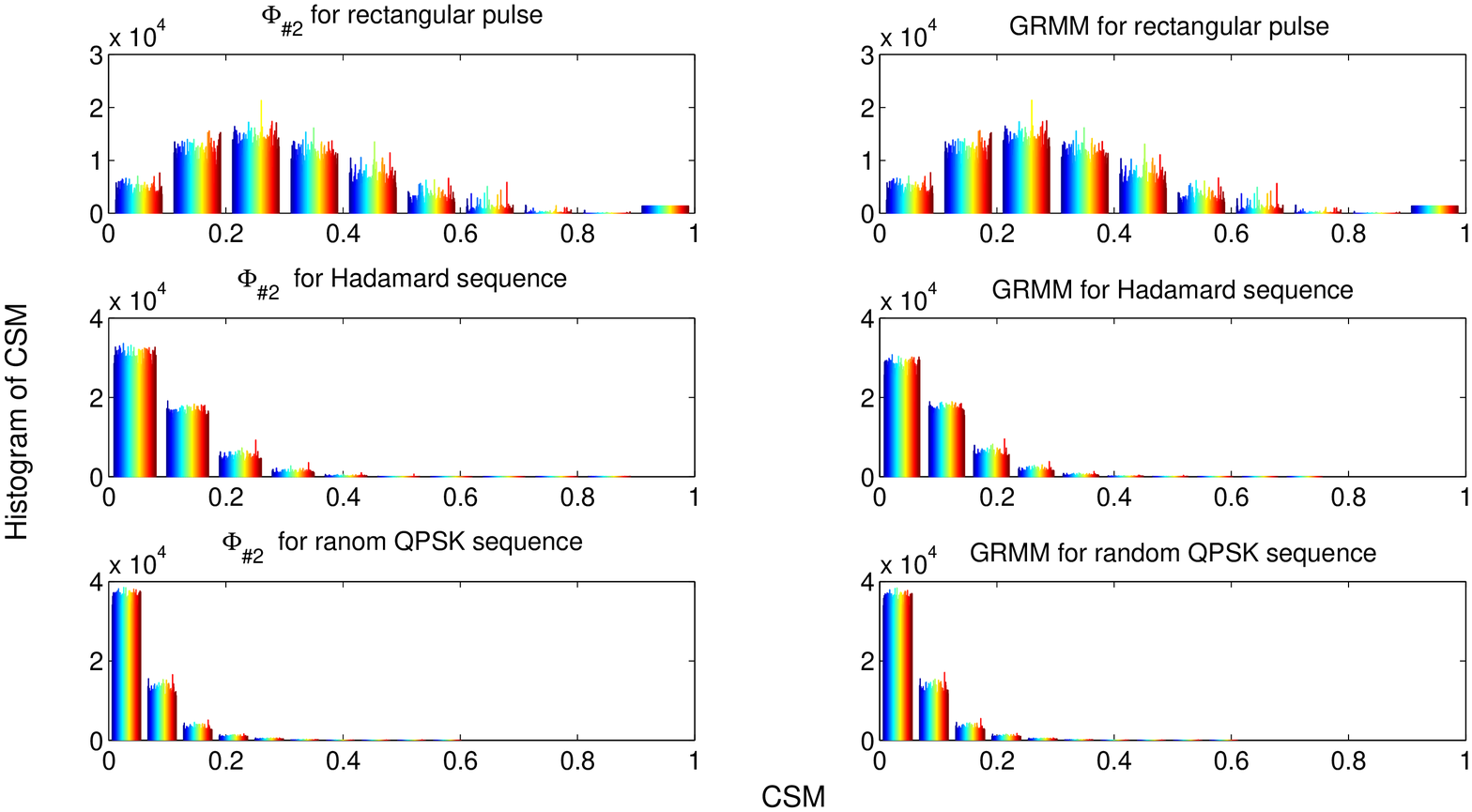,height=3.5in,width=15.0cm}}
\caption{{Coherence of adjacent columns of the sensing
 matrix based on ${\bf \Phi}_{\#2}$
 for different transmit sequences ($M=30$ and $M_t=N_r=10$).
 }}
 \label{correlation_sequence}
\end{figure}

Figure \ref{ConditionaNumber_MaximumCoherence} shows the histograms
of the condition number and the maximum CSM using ${\bf \Phi}_{\#2}$
for Hadamard waveforms and the GRMM produced in 100 random and
independent runs. We consider the case of $M=30$, $M_t=N_r=10$ and
the grid step of the discretized angle-range space is
$[0.5\textordmasculine,15m]$. One can see that the numerical results
fit the observations in Section \ref{coherence_sopt2}, i.e.,  ${\bf
\Phi}_{\#2}$ increases
 the maximum CSM as compared to
the GRMM with high probability. In Fig. \ref{correlation_sequence}
we use histograms to compare the CSM corresponding to adjacent
columns over $100$ independent and random runs.  Although ${\bf
\Phi}_{\#2}$ incorporates information about the waveforms, the
distribution of the column correlation does not change significantly
as compared to that of the conventional matrix. Among the three
types of waveforms, the rectangular pulse gives rise to  the worst
CSM distribution, indicating that the
 performance of the proposed CS approach would  be significantly degraded
if rectangular
 pulses are transmitted. This is because the high autocorrelation of
 the rectangular pulse  results in  high CSM independently of the measurement
matrix used.

%In Table 1, we compare the CSM and the SIR based on
% ${\bf \Phi}_{\#2}$ using rectangular pulse, Hadamard
% waveforms and randomly generated QPSK waveforms. One can see that Hadamard waveforms  can enhance SIR
%and preserve low coherence of the sensing matrix simultaneously.

% \begin{table}
% \caption {The comparison of SIR and CSM}
%\begin{center}
%\begin{tabular}{|c|c|c|c|}
%\hline   &    Rectangular pulse & Hadamard & QPSK
%\\
%\hline    SIR  & best &second best & worst\\
%\hline     Coherence     & worst & second best & best\\
% \hline
%\end{tabular}
%\end{center}
%\label{comparison}
%\end{table}
\subsection{ The  proposed measurement matrix ${\bf \Phi}_{\#1}$ }

We consider a scenario in which  $M_t=N_r=4$ and three stationary
targets exist. The azimuth angle and range of three targets are
randomly generated in 100 runs within
$[0\textordmasculine,1\textordmasculine]$ and $[1000m,1090m]$,
respectively. The data of only one pulse is used and thus only the
angle-range estimates can be obtained. The spacing of adjacent
angle-range grid points is $[0.2\textordmasculine, 15m]$.
$\bf{\Phi}_{\#1}$ is obtained from (\ref{measure_opt_new}) based on
the special structure of (\ref{subopt1}).  $\bf{\Phi}$ in
(\ref{subopt1}) is replaced with $\bf{\Phi}_{\#2}$. We consider
different values of the tradeoff coefficient $\tilde{\lambda}$ in
(\ref{measure_opt_new}).  Transmit nodes send Hadamard waveforms of
length $L=128$. Only $M=20$ measurements per pulse are collected and
forwarded to the fusion center by each node for CS-based MIMO radar
while $100$ measurements are used by the MIMO radar based on the
matched filter method (MFM) \cite{Levanon:04}.

\begin{figure}[htbp]
\centerline{\epsfig{figure=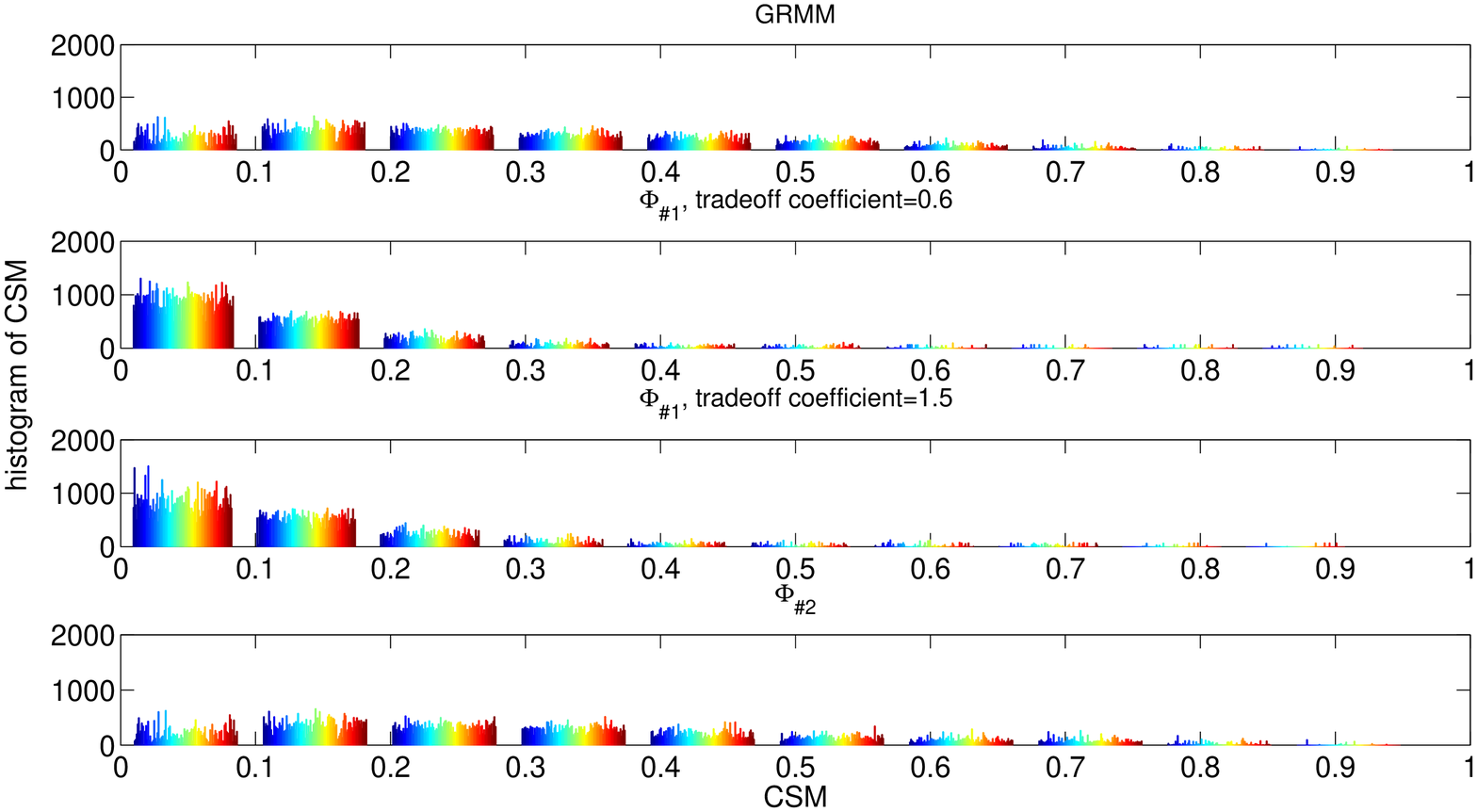,height=3.5in,width=12.0cm}}
\caption{{Coherence distribution of cross columns of the sensing
 matrix using
$\bf{\Phi}_{\#1}$, $\bf{\Phi}_{\#2}$ and the GRMM ($M_t=N_r=4$).
 }}
 \label{distribution_coherence}
 \end{figure}

\begin{figure}[htbp]
 \centerline{\epsfig{figure=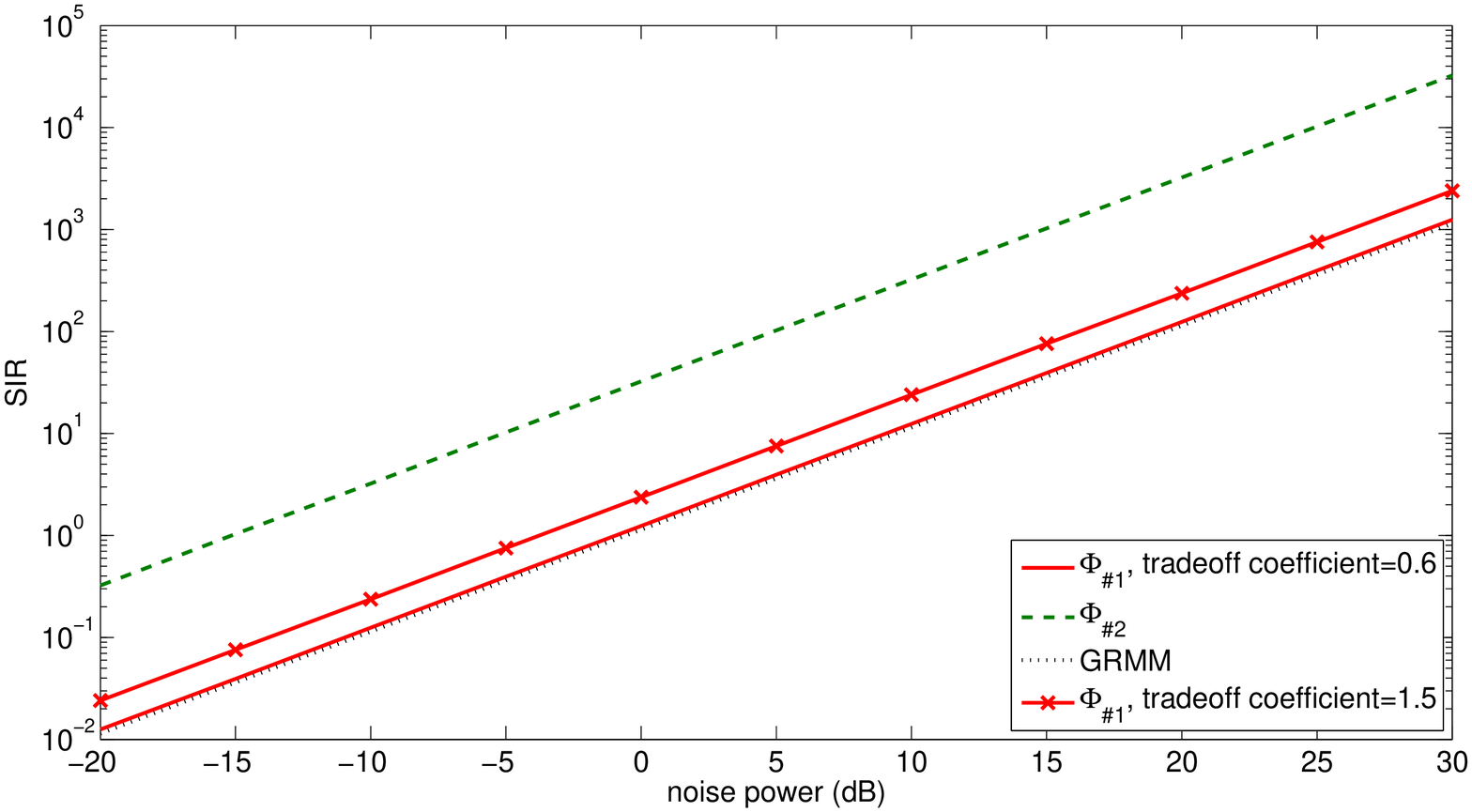,height=3in,width=12.0cm}}
\caption{{SIR for CS-based MIMO radar using $\bf{\Phi}_{\#1}$,
$\bf{\Phi}_{\#2}$ and GRMM for different values of noise power
($M_t=N_r=4$).
 }}
 \label{SIR_all_snr}
 \end{figure}

 Figure \ref{distribution_coherence} shows the
distribution of CSM for the GRMM, $\bf{\Phi}_{\#1}$ and
$\bf{\Phi}_{\#2}$ in 100 random and independent runs.    One can see
that the GRMM and $\bf{\Phi}_{\#2}$ lead to  similar coherence
distributions. $\bf{\Phi}_{\#1}$ slightly reduces the maximum CSM
and significantly increases the number of column pairs with low
coherence as compared to  the other two measurement matrices.
$\bf{\Phi}_{\#1}$ obtained from (\ref{measure_opt2}) using
$\tilde{\lambda}=0.6$ and $\tilde{\lambda}=1.5$ produce a similar
coherence distribution. Figure \ref{SIR_all_snr} shows  the SIR
performance of CS-based MIMO radar using the GRMM, $\bf{\Phi}_{\#1}$
and $\bf{\Phi}_{\#2}$, for different values of noise power in the
absence of a jammer. One can see from Fig. \ref{SIR_all_snr} that
$\bf{\Phi}_{\#2}$ outperforms the other two measurement matrices in
terms of SIR. $\bf{\Phi}_{\#1}$ obtained from (\ref{measure_opt2})
using $\tilde{\lambda}=0.6$ yields slightly better SIR than GRMM. As
expected, increasing $\tilde{\lambda}$ from $0.6$ to $1.5$
moderately improves SIR.

\begin{figure}[htbp]
\centerline{\epsfig{figure=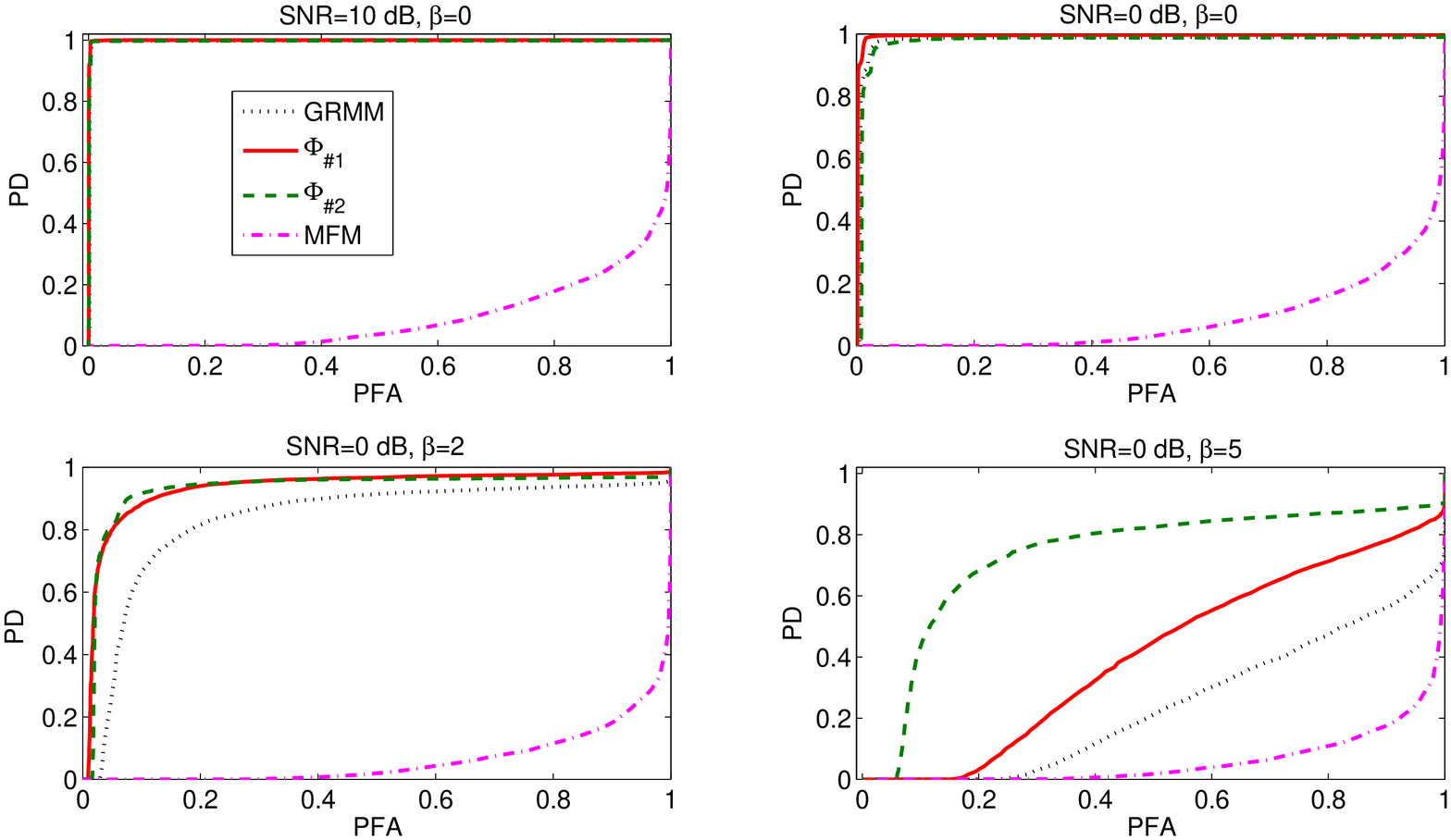,height=4in,width=18.0cm}}
\caption{{ROC curves for CS-based MIMO radar using
$\bf{\Phi}_{\#1}$, $\bf{\Phi}_{\#2}$ and the GRMM and for MIMO
radar using the MFM ($M_t=N_r=4$ and $\tilde{\lambda}=0.6$).
 }}
 \label{ROC1}
\end{figure}

\begin{figure}[htbp]
\centerline{\epsfig{figure=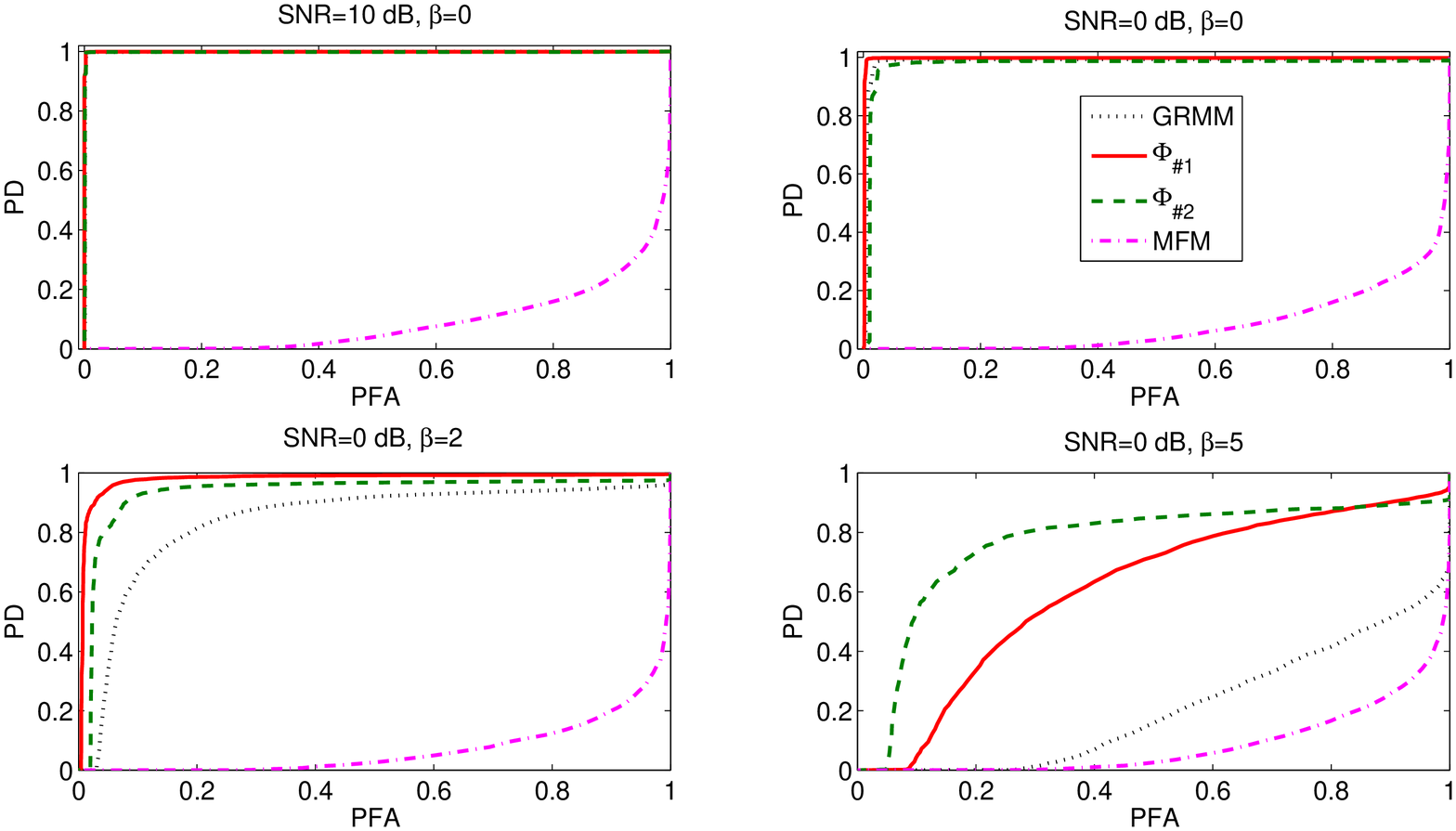,height=4in,width=18.0cm}}
\caption{{ ROC curves for CS-based MIMO radar using
$\bf{\Phi}_{\#1}$, $\bf{\Phi}_{\#2}$ and the GRMM and for MIMO
radar using the MFM ($M_t=N_r=4$ and $\tilde{\lambda}=1.5$).
 }}
 \label{ROC2}
\end{figure}

 Figures \ref{ROC1} and  \ref{ROC2}
compare the ROC performance of CS-based MIMO radar using the three
aforementioned measurement matrices and MIMO radar based on the MFM,
for different combinations of SNR and jammer-signal power. The
probability of detection (PD) here denotes the percentage of cases
in which all the targets are detected. The percentage of cases in
which false targets are detected is denoted  by the probability of
false alarm (PFA). It is demonstrated in Figs. \ref{ROC1} and
\ref{ROC2} that $\bf{\Phi}_{\#1} $ and $\bf{\Phi}_{\#2} $ with
Hadamard waveforms can improve detection accuracy as compared to the
GRMM in the case of mild and strong interference, respectively.
Since an increase in the tradeoff coefficient $\tilde{\lambda}$ can
enhance SIR, $\bf{\Phi}_{\#1} $ obtained from (\ref{measure_opt2})
using $\tilde{\lambda}=1.5$ performs better in the case of strong
interference than using $\tilde{\lambda}=0.6$. Note that the three
measurement matrices give rise to similar performance for $SNR=10
dB$ and $\beta=0$. This is because the interference is sufficiently
small so that all the measurement matrices perform well. Again, one
can see that the MFM is inferior to the CS approach although it uses
far more measurements than the CS approach.

\begin{figure}[htbp]
\centerline{\epsfig{figure=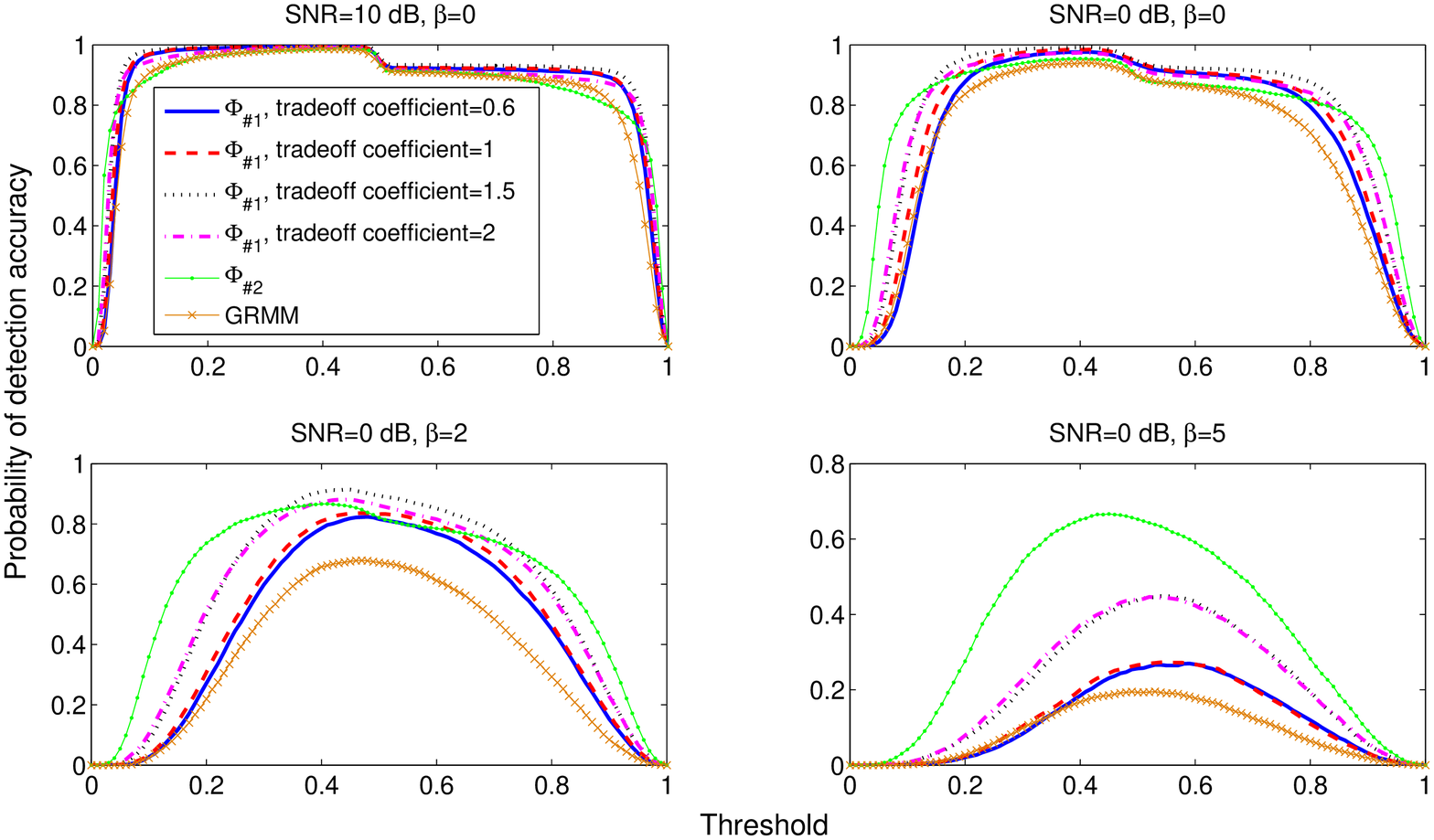,height=4in,width=18.0cm}
} \caption{{ The probability of detection accuracy for CS-based MIMO
radar using $\bf{\Phi}_{\#1}$, $\bf{\Phi}_{\#2}$ and the GRMM for
different values of $\tilde{\lambda}$ ($M_t=N_r=4$).
 }}
 \label{PA}
\end{figure}

It has been seen from Figs. \ref{ROC1} and  \ref{ROC2} that  the
tradeoff coefficient $\tilde{\lambda}$ affects  the performance of
CS-based MIMO radar  using ${\bf \Phi}_{\#1}$. In order to further
investigate the effect of $\tilde{\lambda}$, the curves of
probability of detection accuracy are shown in Fig. \ref{PA} for
$\tilde{\lambda}=0.6,1,1.5,2$ for different thresholds of hard
detection. The probability of detection accuracy here denotes the
percentage of cases in which no real targets are missing and no
false targets exist.
%no false estimation as well.
 By taking all  four combinations of SNR
and jammer-signal power into account, $\tilde{\lambda}=1.5$
%excels the other three values of tradeoff coefficient.
results in the best performance.
 For a particular
case, the optimal tradeoff coefficient depends on multiple factors,
i.e., the basis matrix and the interference. The manner in which SIR
and the CSM affect  the support recovery of a sparse signal still
remains unknown. Therefore, it is impossible to theoretically
determine the optimal tradeoff coefficient.

\section{Conclusions}
We have proposed two measurement matrices in order to improve target
detection performance of CS-based MIMO radar for the case in which
the targets may be located across several range bins. The first one
$\bf{\Phi}_{\#1} $ aims at enhancing SIR and reducing the CSM at the
same time. It is obtained by solving a convex optimization problem.
This measurement matrix requires a heavy computational load as
compared to the conventional measurement matrix, and also needs to
adapt to a particular basis matrix. The computational burden of
solving $\bf{\Phi}_{\#1 }$  can be alleviated through reducing the
number of variables involved in the optimization problem.  The
second proposed measurement matrix $\bf{\Phi}_{\#2} $ targets
improving SIR only. It is constructed based on the transmit
waveforms and also accounts for all possible discretized delays of
target returns within the given time window. $\bf{\Phi}_{\#2 }$  is
dependent on the range grid only and requires much lower complexity
than does $\bf{\Phi}_{\#1} $. It is shown that $\bf{\Phi}_{\#2} $
based on reduced bandwidth transmit waveforms can improve SIR, but
on the other hand, using waveforms that are too narrowband increases
the CSM, thus invalidating conditions for the application of the CS
approach. Therefore,  the waveforms must be chosen carefully to
guarantee the desired performance using the second  measurement
matrix. Numerical results show that $\bf{\Phi}_{\#1} $ and
$\bf{\Phi}_{\#2} $ with the proper waveforms (e.g., Hadamard codes)
can improve detection accuracy as compared to the Gaussian random
measurement matrix in the case of small and strong interference,
respectively.

\bigskip
\centerline{\bf Acknowledgment} The authors would like to thank Dr.
Rabinder Madan of the Office of Naval Research for sharing his ideas
on the use of compressive sampling in the context of MIMO radar.

\bibliographystyle{IEEE}

\begin{thebibliography}{xx}

\bibitem{Fishler:04} E. Fishler, A. Haimovich, R. Blum, D. Chizhik, L.
Cimini and R. Valenzuela,  \newblock ``MIMO radar: An idea whose
time has come," in \newblock {\em   Proc. IEEE Radar Conf.},
Philadelphia, PA, pp. 71-78, Apr. 2004.

\bibitem{Xu:06} L. Xu, J. Li and P. Stoica, \newblock ``Radar imaging via
adaptive MIMO techniques,"
 in \newblock {\em Proc. European Signal Process. Conf.}, Florence, Italy,
Sep. 2006.



\bibitem{Li:07} J. Li, P. Stoica, L. Xu and W. Roberts, \newblock ``On
parameter identifiability of MIMO radar,"
\newblock {\em IEEE Signal Process. Lett.}, vol. 14,  no. 12, pp. 968-971,
Dec. 2007.

\bibitem{Haimovich: 08} A.M. Haimovich,  R.S. Blum and L.J. Cimini,
\newblock ``MIMO
radar with widely separated antennas,"
\newblock {\em IEEE Signal
Process. Magazine}, vol. 25, no. 1,  pp. 116-129,   Jan. 2008.

\bibitem{p1-Godrich-Haimovich-2010} H. Godrich, A.M. Haimovich, and R.S.
Blum, \newblock ``Target localization accuracy gain in MIMO radar
based system," \emph{ IEEE Trans. Info. Theory}, vol.56, no.6,
pp.1-21, Jun. 2010.

\bibitem{p1-He-Blum-2010} Q. He, R.S. Blum, H. Godrich, and A.M.
Haimovich, \newblock ``Target velocity estimation and antenna
placement for MIMO radar with widely separated antennas",
\newblock{\em IEEE Journal of
Selected Topics in Signal Process.}, vol. 4, no. 1, pp. 79-100, Feb.
2010.

\bibitem{Stoica: 07m} P. Stoica and    J. Li,  \newblock ``MIMO radar with
colocated antennas,"
\newblock {\em IEEE Signal
Process. Magazine}, vol. 24,  no. 5,  pp. 106-114,   Sep. 2007.

%\bibitem{Bliss:03} W. Bliss and K. W. Forsythe, \newblock `` Multiple-input
%multiple-output (MIMO) radar and imaging: Degrees
%of freedom and resolution," in \newblock {\em   Proc. 37th IEEE
%Asilomar Conf. Signals, Systems, Computers}, vol.1, pp. 54-59, Nov.
%2003.



%\bibitem{Skolnik:02} M. Skolnik,
%\newblock {\em Introduction to Radar System}, McGraw-
%Hill, 3rd edition, 2002.




\bibitem{Chen:081} C. Chen and P.P. Vaidyanathan, \newblock ``MIMO radar
space-time adaptive processing using prolate spheroidal wave
functions,"
\newblock {\em IEEE Trans. Signal Process.}, vol. 56,  no. 2, pp. 623-635,
Feb. 2008.


%\bibitem{Stoica:07}P. Stoica,    J. Li   and Y. Xie, \newblock ``On probing
%signal
%design for MIMO radar,"  \newblock {\em  IEEE Trans. Signal
%Process.},   vol. 55, pp. 4151-4161, Aug. 2007.

%\bibitem{Aittomaki:07}T. Aittomaki and V. Koivunen, \newblock ``Signal
%covariance matrix
%optimization for transmit beamforming in MIMO radars," in
%\newblock {\em Proc. 38th Asilomar Conf. Signals, Syst. Comput.},  pp.
%182-186, Pacific Grove, CA, Nov. 2007

%\bibitem{Fuhrmann:08}D. R. Fuhrmann and G. San Antonio, \newblock
%``Transmit beamforming for MIMO
%radar systems using signal cross-correlation,"  \newblock {\em  IEEE
%Trans.
% Aerospace and Electronic Systems}, vol. 44, pp. 171 - 186, January 2008.

\bibitem{Donoho:06} D.V. Donoho,   \newblock ``Compressed sensing,"
\newblock {\em IEEE Trans. Info. Theory}, vol. 52, no. 4, pp.
1289-1306,  Apr. 2006.

\bibitem{Candes:061} E.J. Candes, \newblock ``Compressive sampling,"
in \newblock {\em Proc.  The Int'l Congress of Mathematicians},
Madrid, Spain, pp. 1433-1452, Aug. 2006.

\bibitem{Candes:08} E.J. Candes and  M.B. Wakin,  \newblock ``An
introduction to compressive sampling [A sensing/sampling paradigm
that goes against the common knowledge in data acquisition],"
\newblock {\em IEEE Signal
Process. Magazine}, vol. 25, no. 2, pp. 21-30 , Mar. 2008.

\bibitem{Romberg:08} J. Romberg, \newblock ``Imaging via compressive
sampling [Introduction to compressive sampling and recovery via
convex programming],"
\newblock {\em IEEE Signal
Process. Magazine}, vol. 25, no. 2, pp. 14-20, Mar. 2008.

\bibitem{Candes:062} E.J. Candes,  J.K. Romberg and T. Tao,
\newblock ``Stable signal recovery from incomplete and inaccurate
measurements,"
\newblock
{\em Communications on Pure and Applied Mathematics},  vol. 59, no.
8,
 pp. 1207-1223, Aug. 2006.

 \bibitem{Baraniuk:07} R. Baraniuk and P. Steeghs, \newblock ``Compressive
radar imaging,"  in \newblock {\em Proc. IEEE Radar Conf.}, Boston,
MA,  pp. 128-133,  Apr. 2007.

\bibitem{Gurbuz:07} A.C. Gurbuz, J.H. McClellan and W.R. Scott,  \newblock
``Compressive sensing for GPR imaging," in \newblock {\em  Proc.
41th Asilomar Conf. Signals, Syst. Comput.},    Pacific Grove, CA,
pp. 2223-2227, Nov. 2007.

\bibitem{Herman}
M.A. Herman and T. Strohmer, \newblock``High-resolution radar via
compressed sensing,''  {\it IEEE Trans. Signal Process.}, vol. 57,
no. 6, pp. 2275-2284,  Jun. 2009.

\bibitem{Herman:08}  M. Herman and T. Strohmer, \newblock ``Compressed
rensing radar," in \newblock {\em Proc. IEEE Int'l Conf.  Acoust.
Speech Signal Process}, Las Vegas, NV,  pp. 2617-2620,  Mar.-Apr.
2008.

\bibitem{Petropulu:08} A.P. Petropulu, Y. Yu and H.V. Poor,  \newblock
``Distributed MIMO radar using compressive sampling," in \newblock
{\em  Proc. 42nd Asilomar Conf. Signals, Syst. Comput.}, Pacific
Grove, CA, pp. 203-207, Nov. 2008.

\bibitem{Chen:08} C.Y. Chen and P.P. Vaidyanathan,  \newblock ``Compressed
sensing in MIMO radar," in \newblock {\em  Proc. 42nd Asilomar Conf.
Signals, Syst. Comput.},  Pacific Grove, CA, pp. 41-44, Nov. 2008.

\bibitem{Strohmer:Asilomar09}  T. Strohmer and B. Friedlander,  \newblock
``Compressed sensing for MIMO radar - algorithms and performance,"
in \newblock {\em  Proc. 43rd Asilomar Conf. Signals, Syst.
Comput.},  Pacific Grove, CA, pp. 464-468, Nov. 2009.


\bibitem{Yu:09_tsp}  Y. Yu, A.P. Petropulu and H.V. Poor, \newblock ``MIMO
radar using compressive sampling," \newblock {\em IEEE Journal of
Selected Topics in Signal Process.}, vol. 4, no. 1, pp. 146-163,
Feb. 2010.











\bibitem{Yu:10_aes} Y. Yu, A.P. Petropulu and H.V. Poor, \newblock ``CSSF
MIMO radar: Low-complexity   compressive sensing based MIMO radar
that uses step frequency," submitted to \newblock {\em IEEE Trans.
Aerospace and Electronic Systs.}.






\bibitem{Candes:07} E.J. Candes and T. Tao, \newblock ``The Dantzig
selector: Statistical estimation when $p$ is much larger than $n$,"
\newblock {\em  Ann. Statist.}, vol. 35, no. 6, pp. 2313-2351, Dec.
2007.

\bibitem{Dai:09} W. Dai and O. Milenkovic, \newblock ``Subspace pursuit for
compressive sensing signal reconstruction," \emph{IEEE Trans. Info.
Theory}, vol. 55, no. 5, pp. 2230-2249, May 2009.

\bibitem{Needell:09} D. Needell and J. Tropp, \newblock ``CoSaMP: Iterative
signal recovery from incomplete and inaccurate samples," \emph{Appl.
Comput. Harmonic Anal.}, vol. 26, no. 3, pp. 301-321, May 2009.

\bibitem{Tang:09} G. Tang and A. Nehorai, \newblock ``Performance analysis
for sparse support recovery," \emph{IEEE Trans. Info. Theory}, vol.
56, no. 3, pp. 1383-1399, Mar. 2009.

%\bibitem{Chen:Asilomar08} C. Y. Chen and P. P. Vaidyanathan,  \newblock ``Compressed sensing in MIMO radar,"
%\newblock {\em  Proc. 42nd Asilomar Conf. Signals, Syst. Comput},  Pacific Grove, CA, Nov.
%2008.

\bibitem{Tropp:04} J.A. Tropp, \newblock ``Greed is good: Algorithmic
results for sparse approximation," \emph{IEEE Trans. Info. Theory},
vol. 50, no. 10, pp. 2231-2242, Oct. 2004.

\bibitem{Tropp:09} J.A. Tropp, \newblock ``Just relax: Convex programming
methods for identifying sparse signals," \emph{IEEE Trans. Info.
Theory}, vol. 55, no. 2, pp. 917-918, Feb. 2009.




























\bibitem{Ochiai}
H. Ochiai, P. Mitran, H.V. Poor and V. Tarokh,
\newblock``Collaborative beamforming for distributed wireless ad hoc
sensor networks,'' {\it IEEE Trans. Signal Process.}, vol. 53,  no.
11, pp. 4110-4124, Nov. 2005.

%\bibitem{Krim:2006}
%H. Krim and M. Viberg, \newblock``Two decades of array signal
%processing research: The parametric approach," {\it IEEE Signal
%Processing Magazine}, vol. 13,   pp. 67 - 94, July 1996.



\bibitem{Levanon:04} N. Levanon and E. Mozeson, \emph{Radar
Signals},
 Hoboken, NJ: J. Wiley, 2004.


















%\bibitem{Tropp:07} J.A. Tropp and A.C. Gilbert,  \newblock ``Signal
%recovery from random
%measurements via orthogonal matching pursuit,"  \newblock {\em IEEE
%Trans. on Information Theory}, vol. 53, no.12, pp. 4655-4666, 2007.






%\bibitem{Mallat:1993}S. G. Mallat and Z. Zhang, "Matching Pursuits with
%Time-Frequency
%Dictionaries,'' {\it IEEE Transactions on Signal Processing},  pp.
%3397-3415, December 1993.
%
%\bibitem{Chen:1994}Shaobing Chen and D.V. Donoho,  "Basis pursuit,'' {\it
%Proc. 28th Asilomar Conf. Signals, Syst. Comput}, vol. 1, pp.
%41-44, 31 Oct.-2 Nov. 1994.





%\bibitem{Boyd:04}
%S. Boyd and L. Vandenberghe, {\it Convex Optimization.} Cambridge
%University Press, Cambridge, UK, 2004.
\end{thebibliography}

\appendices

\end{document}